\documentclass[11pt,amsmath,amssymb,nofootinbib,notitlepage,longbibliography]{revtex4-1}
\usepackage{amsmath,amssymb}
\usepackage{slashed}
\usepackage{graphicx}
\usepackage{bm}
\usepackage[top=1.0in,bottom=1.0in,left=1.0in,right=1.0in]{geometry}
\usepackage[colorlinks,linkcolor=blue,citecolor=blue]{hyperref}

\newcommand{\be}{\begin{equation}}
\newcommand{\ee}{\end{equation}}
\newcommand{\bea}{\begin{eqnarray}}
\newcommand{\eea}{\end{eqnarray}}
\newcommand{\ba}{\begin{eqnarray}}
\newcommand{\ea}{\end{eqnarray}}

\begin{document}

\title{Holographic charm and bottom pentaquarks II\\
Open and hidden decay widths}


\author{Yizhuang Liu}
\email{yizhuang.liu@uj.edu.pl}
\affiliation{Institute of Theoretical Physics,
Jagiellonian University, 30-348 Kraków, Poland}

\author{Maciej A. Nowak}
\email{maciej.a.nowak@uj.edu.pl}
\affiliation{Institute of Theoretical Physics and Mark Kac Center for Complex Systems Research,
Jagiellonian University, 30-348 Kraków, Poland}

\author{Ismail Zahed}
\email{ismail.zahed@stonybrook.edu}
\affiliation{Center for Nuclear Theory, Department of Physics and Astronomy, Stony Brook University, Stony Brook, New York 11794--3800, USA}



\begin{abstract}
We analyze the decay modes of the three  $[\frac 12\frac 12^-]_{S=0,1}$  and $[\frac 12\frac 32^-]_{S=1}$ non-strange pentaquarks with hidden charm and bottom,
predicted by holographic QCD   in the heavy quark limit. In leading order, the pentaquarks
 are degenerate and stable by heavy quark symmetry. At next to leading order,  the spin interactions lift the
 degeneracy  and cause the pentaquarks to decay. We show that the open charm (bottom) decay modes dwarf the hidden charm (bottom) ones,
 with total widths that are consistent with those recently reported by LHCb for charm pentaquarks. Predictions for bottom pentaquarks are given.
\end{abstract}

\maketitle

\section{Introduction}

\begin{figure}[!htb]
\includegraphics[height=9cm,width=12cm]{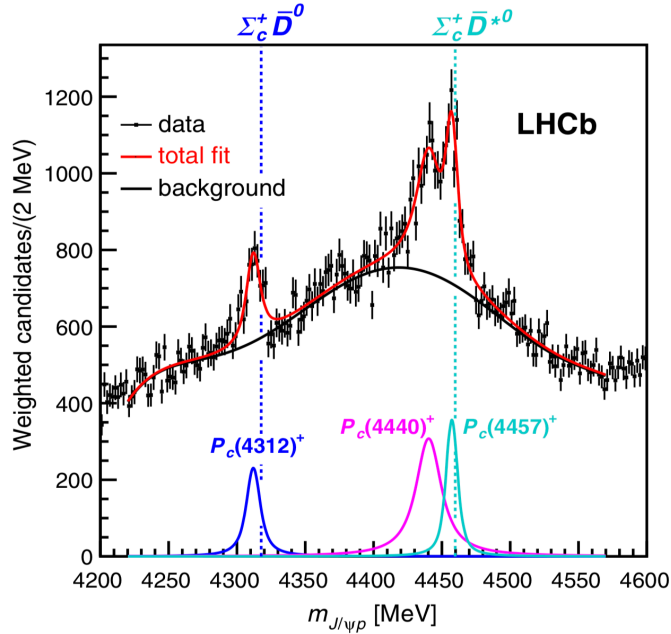}
 \caption{LHCb measurements of the $P_c$ states fitted with three BW distributions red-solid curve and fitted background black-solid curve,
with the  mass thresholds for the $\Sigma_c^+\bar D^0$ and $\Sigma_c^+\bar D^{*0}$ final states  shown for comparison  from~\cite{Aaij:2019vzc}.
}
  \label{fig_ajit}
\end{figure}

The LHCb
high statistics analysis~\cite{Aaij:2019vzc} shows that the previously reported $P_c^+(4450)$~\cite{Aaij:2015tga}
splits into two narrow peaks  $P_c^+(4440)$ and  $P_c^+(4457)$ just below the $\Sigma_c^+\bar D^{*0}$ treshold, with the appearance of a new and
narrow $P_c^+(4312)$ state right below the $\Sigma_c^+\bar D^{0}$. The evidence  for the old  and broad $P_c^+(4380)$~\cite{Aaij:2015tga} has  now
weakened. The  reported charm pentaquark widths are narrow~\cite{Aaij:2019vzc}
\bea
\label{LHCBMW}
m_{P_c}&=&4311.9\pm 0.7\,{\rm MeV}\qquad \Gamma_{P_c}=9.8\pm 2.7\,{\rm MeV}\nonumber\\
m_{P_c}&=&4440.3\pm 1.3\,{\rm MeV}\qquad \Gamma_{P_c}=20.6\pm 4.9\,{\rm MeV}\nonumber\\
m_{P_c}&=&4457.3\pm 0.6\,{\rm MeV}\qquad \Gamma_{P_c}=6.4\pm 2.0\,{\rm MeV}
\eea
The  $P_c(4312)$ is observed to be 10 MeV below the $\Sigma^+_c\bar D^0$ treshold,  and the $P_c(4457)$ just
 5 MeV below the $\Sigma_c^+\bar D^*$ treshold as illustrated  in Fig.~\ref{fig_ajit} from~\cite{Aaij:2019vzc}, a strong indication to their molecular
 origin as discussed
 by many~\cite{Burns:2015dwa,Richard:2016eis,Lebed:2016hpi,Esposito:2016noz,Olsen:2017bmm,Guo:2017jvc,Karliner:2017qhf,Du:2021fmf}
 (and references therein).

In  the heavy quark limit,  the heavy-light pair $[0^-,1^-]=[D, D^*]$ is degenerate and the $\Sigma^+_c\bar D^0$ and
 $\Sigma_c^+\bar D^*$ thresholds coalesce. As a result,  the three reported pentaquark  states become degenerate  and stable by heavy quark symmetry.
Three degenerate and stable pentaquark states  with isospin-spin-parity assignments  $[\frac 12\frac 12^-]_{S=0,1}$  and $[\frac 12\frac 32^-]_{S=1}$,
were predicted by holographic QCD, in the triple limit of a large number of colors,  large $^\prime$t Hooft gauge coupling $\lambda$ and a heavy
quark mass~\cite{Liu:2017xzo,Liu:2017frj}. The same assignments were subsequently
 made using the molecular construction~\cite{Liu:2019tjn,Xiao:2019aya,Du:2019pij,Yan:2021nio}.

The newly reported $P_c(4337)$ with a width of 29 MeV  at  3-sigma significance~\cite{LHCb:2021chn},
 appears to overlap with the reported $P_c(4312)$  at 7-sigma significance, and is not supported by our  holographic analysis of the low-lying
 pentaquark states. The excited even  and  odd  parity  holographic pentaquark states $P_c^*$ lie higher in mass, and are likely much broader by phase space~\cite{Liu:2017xzo,Liu:2017frj,Liu:2021tpq}.

Holographic pentaquarks are composed of heavy-light mesons bound to a topological  instanton core in bulk.
They are the dual of a nucleon core bound to heavy-light mesons at the boundary. In the heavy quark limit, the pentaquarks with hidden
charm and bottom are degenerate, heavy and stable~\cite{Liu:2017xzo,Liu:2017frj,Li:2017dml,Fujii:2020jre}. Away from the heavy
quark limit, spin-spin and spin-orbit forces lift the degeneracy and cause them to decay as we will show below.
This work is a follow up on our recent re-analysis of the charm and bottom pentaquark states including the spin
effects, to which we refer for completeness~\cite{Liu:2021tpq}.



The organization of the paper is as follows:
In section~\ref{ADS} we briefly review  the essential aspects of the holographic construction in leading order in the
heavy quark mass. In section~\ref{OPEN}, we detail the spin contributions to order $1/m_H$ which are at
the origin of the two-body decay of the pentaquarks with open charm final states. In section~\ref{HIDDEN}
we show how the  two-body decay channel with hidden charm can be extracted from a Witten diagram in bulk.
We derive a number of model independent ratios for the decay modes for both charm and bottom pentaquarks.
For charm pentaquarks,
they compare well to the total widths recently reported by LHCb.
Our conclusions are in section~\ref{CONCLUSION}. We include complimentary Appendices for completeness.

\section{ Holographic heavy-light effective action}~\label{ADS}

The D4-D8-D$\bar 8$ set-up for light flavor branes is standard~\cite{Sakai:2004cn}.  The minimal modification that accommodates
heavy mesons makes use of an extra heavy brane as discussed in~\cite{Liu:2016iqo,Liu:2017xzo}.
The effective action  consists of the non-Abelian Dirac-Born-Infeld (DBI), Chern-Simons (CS)  and mass term

\bea
\label{1}
S_{\rm DBI}\approx -\kappa\int d^4x dz\,{\rm Tr}\left({\bf f}(z){\bf F}_{\mu\nu}{\bf F}^{\mu\nu}+{\bf g}(z){\bf F}_{\mu z}{\bf F}^{\nu z}\right)
-\frac 12 m_H^2 \int d^4x dz\,{\rm Tr}\left(\Phi^\dagger_M \Phi_M\right)
\eea
The warping factors are

\be
{\bf f}(z)=\frac{R^3}{4U_z}\,,\qquad {\bf g}(z)=\frac{9}{8}\frac{U_z^3}{U_{KK}}
\ee
with $U_z^3=U_{KK}^3+U_{KK}z^2$ and $\kappa\equiv a\lambda N_c$ and
$a=1/(216\pi^3)$  in units of $M_{KK}$~\cite{Sakai:2004cn}. Our conventions are $(-1,1,1,1,1)$ with $A_{M}^{\dagger}=-A_M$ and
the labels $M,N$ running over $\mu,z$ only in this section.
The effective fields in the field strengths are~\cite{Liu:2016iqo,Liu:2017xzo}

\bea
\label{2}
&&{\bf F}_{MN}=\nonumber \\
&&\left(\begin{array}{cc}
F_{MN}-\Phi_{[M}\Phi_{N]}^{\dagger}&\partial_{[M}\Phi_{N]}+A_{[M}\Phi_{N]}\\
-\partial_{[M}\Phi^{\dagger}_{N]}-\Phi^{\dagger}_{[M}A_{N]}&-\Phi^{\dagger}_{[M}\Phi_{N]}
\end{array}\right)
\eea
The  matrix valued 1-form gauge field is

\be
\label{7}
{\bf A}=\left(\begin{array}{cc}
A&\Phi\\
-\Phi^{\dagger}&0
\end{array}\right)
\ee
For $N_f=2$, the naive Chern-Simons 5-form is

\be
\label{CSNAIVE}
S_{CS}=\frac{iN_c}{24\pi^2}\int_{M_5}\,{\rm Tr}\left(AF^2-\frac{1}{2}A^3F+\frac{1}{10}A^5\right)
\ee
For $N_f$ coincidental branes, the $\Phi$ multiplet is massless, but for separated branes
 they are massive with  $m_H$ fixed by the separation between the heavy and light branes.
 We follow~\cite{Liu:2016iqo}  and fix it by the heavy meson masses
 $M_{D}=1870$ MeV (charmed) and $M_B=5279$ MeV (bottomed)
using

\be
\label{MDMH}
M_{D,B}=m_{H}+\frac{M_{KK}}{2\sqrt{2}}
\ee




In the coincidental brane limit,
light baryons are interchangeably described as a flavor instanton or a D4 brane wrapping the $S^4$.
The instanton size is small with $\rho\sim 1/\sqrt{\lambda}$ after balancing  the order $\lambda$
bulk gravitational attraction  with the subleading and of order $\lambda^0$ U(1) induced topological
repulsion~\cite{Sakai:2004cn}. The O(4) instanton gauge field is of the form

\be
\label{XS3}
A_{M}(y)=-\overline{\sigma}_{MN}\partial_NF(y)\qquad \left.F_{zm}(y)\right|_{|y|=R}=0
\ee
Since $\rho\sim 1/\sqrt{\lambda}$
is the  typical  instanton size, it is convenient to rescale the fields

\bea
\label{S3}
(x_0, x_{M})\rightarrow (x_0,x_{M}/\sqrt{\lambda}), \sqrt{\lambda}\rho\rightarrow \rho\qquad\qquad
(A_{0},A_M)\rightarrow (A_0, \sqrt{\lambda}A_M)
\eea
with the leading order equations of motion

\be
\label{HL1}
D_{M}F_{MN}=0\qquad \partial_M^2A_0=-\frac 1{32\pi^2 a}F_{aMN}\star {F}_{aMN}
\ee
Similarly, the bound heavy-light fields  $(\Phi_0, \Phi_M)$ are rescaled using

\be
\label{S3S}
(\Phi_0,\Phi_M)\rightarrow (\Phi_0, \sqrt{\lambda}\Phi_M)
\ee

Following the rescaling, the effective action for  the light gauge fields $(A_0, A_M)$ and
the heavy fields $(\Phi_0,\Phi_M)$ in leading order  is~\cite{Liu:2016iqo,Liu:2017xzo}

\be
\label{RS1}
{\cal L}=aN_c\lambda {\cal L}_{0}+aN_c{\cal L}_{1}+{\cal L}_{CS}
\ee
with

\bea
\label{ACTIONALL}
{\cal L}=aN_c\lambda {\cal L}_0+aN_c({\cal L}_1+\tilde {\cal L}_1)+{\cal L}_{\rm CS}
\eea
with each contribution given by

\bea
{\cal L}_0=&&-(D_M\Phi_N^{\dagger}-D_N\Phi_M^{\dagger})(D_M\Phi_N-D_N\Phi_M)+2\Phi_M^{\dagger}F_{MN}\Phi_N  \ ,\nonumber \\
{\cal L}_1=&&+2(D_0\Phi_M^{\dagger}-D_M\Phi_0^{\dagger})(D_0\Phi_M-D_M\Phi_0)-2\Phi_0^{\dagger}F^{0M}\Phi_M\nonumber\\
&&-2\Phi^{\dagger}_MF^{M0}\Phi_0 -2m_H^2\Phi^{\dagger}_M\Phi_M \ ,
\nonumber\\
\tilde {\cal L}_1=&&+\frac{z^2}{3}(D_i\Phi_j-D_j\Phi_i)^{\dagger}(D_i\Phi_j-D_j\Phi_i) \nonumber \\
&&-2z^2(D_i\Phi_z-D_z\Phi_i)^{\dagger}(D_i\Phi_z-D_z\Phi_i)
-\frac{2}{3}z^2\Phi_i^{\dagger}F_{ij}\Phi_j+2z^2(\Phi^{\dagger}_zF_{zi}\Phi_i+c.c) \ \nonumber\\
{\cal L}_{CS}=&&-\frac{iN_c}{16\pi^2}\Phi^{\dagger}(dA+A^2)D\Phi-\frac{iN_c}{16\pi^2}(D\Phi)^{\dagger}(dA+A^2) \Phi +{\cal O}(\Phi^3)\ .
\eea
The expansion around the heavy quark limit will be sought using
$\Phi_{M}=\phi_{M}e^{-im_Hx_0}$ for particles and $m_H\rightarrow -m_H$ for anti-particles.
In particular, we have in leading order~\cite{Liu:2017xzo,Liu:2017frj}

\be
\label{RX66X}
{\cal L}_0=-\frac 12 \left|f_{MN}-\star f_{MN}\right|^2+2\phi_M^\dagger (F_{MN}-\star F_{MN})\phi_N
\ee
subject to  the constraint equation $D_M\phi_M=0$ with $f_{MN}=\partial_{[M}\phi_{N]}+A_{[M}\phi_{N]}$, and

\bea
\label{RX5}
\frac{{\cal L}_{1}}{aN_c}\rightarrow 4m_H\phi^{\dagger}_{M}iD_0\phi_{M}\qquad\qquad
{\cal L}_{CS}\rightarrow \frac{m_H N_c}{16\pi^2}\phi^{\dagger}_{M}\star F_{MN}\phi_{N}
\eea
For self-dual fields $F_{MN}=\star F_{MN}$,  and the minimum of (\ref{RX66X})  is reached for
$f_{MN}=\star f_{MN}$. As a result,  the combination $\psi=\bar \sigma_{M}\phi_{M}$
with $\sigma_M =(i, \vec \sigma)$, obeys the zero mode equation $\sigma_{M}D_{M}\psi= D \psi =0$.
While binding to the core instanton, the heavy mesons with spin-1 transmute to a Weyl fermion
with spin-$\frac 12$~\cite{Liu:2017xzo,Liu:2017frj}.

The holographic charmed pentaquark states are ultimatly bound topological molecules  with hidden charm, without the ambiguities
related to the type of meson exchange to use and the details of the form factors (hard core),  a challenge for most molecular
constructions~\cite{Burns:2015dwa,Richard:2016eis,Lebed:2016hpi,Esposito:2016noz,Olsen:2017bmm,Guo:2017jvc,Karliner:2017qhf,Eides:2017xnt,Lin:2019qiv,Du:2021fmf}
(and references therein).  The dual of the hard core is  the instanton core which is universal and fixed by gauge-gravity interactions in bulk.
The dual of the meson exchanges are bulk  light and heavy gauge fields
regulated by unique D-brane gauge interactions  in conformity with chiral symmetry, vector dominance and heavy quark symmetry at the boundary.
 We now address their strong decay modes using  the effective action (\ref{ACTIONALL}).

\section{Open charm decays}~\label{OPEN}

The charmed pentaquark states decay modes can proceed through either open charm channels given their
proximity to the  $\Sigma_c[\bar D,\bar D^*]$ thresholds~\cite{Aaij:2019vzc}, or hidden charm channel such as
$J/\Psi$ as originally observed~\cite{Aaij:2015tga}. For clarity, all the analyses to follow will
be carried with the decay kinematics  using $P_c(4440)$.  The final results will be tabulated for all three
charm pentaquark states recently reported, and extended to the yet to be observed bottom pentaquarks.


The decay modes follow from the coupling between the background classical field $\Phi_M$ sourced by the baryonic moduli,
and the fluctuating heavy-light meson field $\delta \Phi_M$~\cite{Liu:2016iqo}. Note that our classical field configuration $(\Phi_0,\Phi_M)$ only solves the equation of motion to leading order in $1/\lambda$. Therefore  under the shift
$\Phi_M\rightarrow\delta \Phi_M$ there are linear terms in $\delta \Phi_M$. They do not affect the stability
of the instanton core.

More specifically, the linear contributions in leading order  in $m_H$ are

\begin{align}
\delta {\cal L}=4iaN_cm_H\left(\delta \Phi_M^{\dagger}\hat A_0 \Phi_M +\Phi_M^{\dagger}\hat A_0 \delta \Phi_M\right)+\frac{N_cm_H}{8\pi^2}\left(\delta \Phi_M^{\dagger}F_{MN}\Phi_N+ \Phi_M^{\dagger}F_{MN}\delta \Phi_N\right) \ ,
\end{align}
The first contribution is  kinetic and the second contribution is topological
(Chern-Simons term).  For vector mesons  we have

\begin{align}
\delta \Phi_M(t,\vec{x},z)=\epsilon_{M}e^{-iM_nt}\phi_n(z) \ , \qquad\qquad
\delta \Phi_z(t,\vec{x},z)=0 \ ,
\end{align}
with the interaction term

\begin{align}\label{eq:changeL}
\delta L=\frac{im_HN_c}{2\pi^2\sqrt{16m_HaN_c}}\frac{c\phi_n(Z)}{(X^2+\rho^2)^{\frac{5}{2}}}\left(1+\frac{5\rho^2}{2(X^2+\rho^2)}\right)\left(\vec{\epsilon}^{\dagger}\cdot \vec{\sigma}\chi_{Q}-\chi^{\dagger}_{Q}\vec{\sigma}\cdot \vec{\epsilon}\right)  \ ,
\end{align}
The heavy-light mesonic wavefunctions in bulk satisfy ($\tilde Z=\sqrt{m_H}Z$)

\begin{align}
-\frac{d^2\phi_n(\tilde Z)}{d\tilde Z^2} +\frac{\tilde Z^2}{2} \phi_n(\tilde Z)=(m_n^2-m_H^2)\phi_n(\tilde Z) \ ,
\end{align}
with  the  normalized solutions~\cite{Liu:2016iqo}

\begin{align}
\phi_n( \tilde Z)=\frac{1}{\sqrt{2\tilde\kappa}}\frac{1}{\sqrt{2^n n!}}\bigg(\frac{\sqrt{2}}{2\pi}m_H\bigg)^{\frac{1}{4}}e^{-\frac{\sqrt{2}\tilde Z^2}{4}}
H_n\left[\bigg(\frac{\sqrt{2}}{2}\bigg)^{\frac{1}{2}}\tilde Z\right] \ .
\end{align}
and the Reggeized mass spectrum

\be
m_n^2\approx m_H^2+\frac {7m_Hm_\rho}4\bigg(n+\frac 12\bigg)
\ee
Note that the two brane tensions $\tilde\kappa$ in the heavy-light sector  and $\kappa$ in the light-light  sector are identified,
for bulk filling branes.
However here, we will keep them separate phenomenologically as we discuss below. With this in mind, the Hamiltonian
following from (\ref{eq:changeL}) after integration over $dZd^3X$, reads

\begin{align}
\label{HAMX}
\delta H= i\alpha\epsilon_i^{\star}\tau_i\lambda-i\alpha\lambda^{\dagger}\tau_i\epsilon_i+\alpha \epsilon^{\dagger}\lambda+\alpha \lambda^{\dagger}\epsilon
\end{align}
with  the moduli coefficient

\begin{align}
\label{COEF2}
\alpha\bigg(\rho,\frac{Z}{\rho}\bigg)=\frac{\sqrt{2}\rho N_c}{3\pi^2\sqrt{aN_c}(Z^2+\rho^2)}\int dZ \phi_n(Z) \ ,
\end{align}
which depend on the specifics of the  moduli wavefunctions which are detailed in Appendix~\ref{COEFFICIENT}.

\subsection{Generic  form of the spin interaction}

If we fix the vector meson polarization to say $\epsilon_M$, then the sole coupling to the angular momentum is rotor-like
$\chi^a F^a(\chi)$  which contributes to the Hamiltonian as $ \vec{L}\cdot \vec{F} $.
It  {\it conserves}  angular momentum $l$ and cannot cause an angular momentum transition necessary for the open channel decays.
However, a close inspection shows that we need to consider  the mismatch caused by the gauge-transformation
$V$ that acts solely  on the instanton-profile but not on the external field~\cite{Hashimoto:2008zw}. Including this gauge-transformation amounts to the substitution

\begin{align}
\delta \Phi_M \rightarrow V^{-1}(t,z,x)\delta \Phi_M \approx (a_4- i\vec{a}\cdot \vec{\tau}) \delta \Phi_M \ .
\end{align}
As a result, the change in the Lagrangian can still be obtained from equation (\ref{eq:changeL}) with the replacement
\begin{align}
\epsilon_M \rightarrow (a_4- i\vec{a}\cdot \vec{\tau})\epsilon_M \ ,
\end{align}
which allows for the transition from $l$ to $l \pm 1$.  We conclude that by expanding in linear order in
$\delta \Phi_M$, we can generate a transition vertex  with net  angular momentum change by $1$.   Therefore the transitions
from $P_c$ with $l=1$ to $\Lambda_c$ with $l=0$ and to $\Sigma_c$ with $l=2$ are all possible.

\subsection{General transition vertices}

The  vertex responsible for the decay to a vector meson $P_c\rightarrow D^\star p$ follows from

\begin{align}
\delta H=i\alpha\epsilon_i^{\star}(a_4+i\vec{a}\cdot\tau)\tau_i\lambda-i\alpha\lambda^{\dagger}\tau_i(a_4-i\vec{a}\cdot \tau)\epsilon_i  \ ,
\end{align}
or more specifically the matrix element

\begin{align}
\langle l'm';\frac{1}{2}s|\delta H|lm_1;Sm_2\rangle \ .
\end{align}
\\
\\
{\bf 1.} For $S=0$:
\\
the transition matrix for $P_c\rightarrow D+p$ with a scalar meson final state is

\begin{align}
\delta H=\alpha \epsilon^{\dagger}(a_4+i\vec{a}\cdot \vec{\tau})\lambda+\alpha \lambda^{\dagger}(a_4-i\vec{a}\cdot \vec{\tau})\epsilon \ ,
\end{align}
and the corresponding transition amplitude is

\begin{align}
{\cal M}(P_c,S=0 \rightarrow D(\epsilon)+l'm'+\lambda_s)=\alpha \epsilon^{\dagger}\langle l'm'|a_4+i\vec{a}\cdot \vec{\tau}|lm\rangle \sigma_2 \lambda_s \ .
\end{align}
The transition amplitude with a vector meson final state is

\begin{align}
&{\cal M}(P_c,S=0 \rightarrow D^{\star}(\vec{\epsilon})+l'm'+\lambda_s)\nonumber \\
&=\alpha\langle l'm' |ia_4|lm \rangle \vec{\epsilon}^{\star}\cdot \vec{\tau} \sigma_2 \lambda_s^{\star}-\alpha\langle l'm' |ia_i|lm \rangle(\vec{\epsilon}^{\star}\times \vec{\tau})_i\sigma_2 \lambda_s^{\star}-\alpha\langle l'm'| ia_i|lm\rangle \epsilon^{\star}_i\sigma_2  \lambda_s \ .
\end{align}
\\
\\
{\bf 2.} For $S=1$:
\\
 the transition amplitude for $P_c\rightarrow D+p$ with a scalar meson final state is

\begin{align}
{\cal M}(P_c,1S \rightarrow D(\epsilon)+l'm'+\lambda_s)=\alpha\bigg(\epsilon^{\dagger}\langle l'm'|a_4+i\vec{a}\cdot \vec{\tau}|lm\rangle\bigg)_{s'} C^{1;S}_{s's} \lambda_s \ .
\end{align}
  after replacing  $\sigma_2$ by the general Clebsch-Gordan coefficient for $\frac{1}{2}+\frac{1}{2}=1$

\begin{align}
\sigma_2\rightarrow C^{1;s+s'}_{ss'}=\delta_{ss'}+\sigma^1_{ss'} \ ,
\end{align}
The corresponding transition amplitude with a vector meson final state is

\begin{align}
&{\cal M}(P_c,1S \rightarrow D^{\star}(\vec{\epsilon})+l'm'+\lambda_s)\nonumber \\&=\alpha\langle l'm' |ia_4|lm \rangle (\vec{\epsilon}^{\star}\cdot \vec{\tau})_{s'}C^{1;S}_{s's} \lambda_s^{\star}-\alpha\langle l'm' |ia_i|lm \rangle(\vec{\epsilon}^{\star}\times \vec{\tau})_{is'}C^{1;S}_{s's} \lambda_s^{\star}-\alpha\langle l'm'| ia_i|lm\rangle \epsilon^{\star}_{is'}C^{1;S}_{s's}\lambda^{\star}_s \ .
\end{align}
\\
\\
In a typical decay, we need  to combine $\lambda_s$ with $l'm'$ to form the finite $J$ final state,
 and combine $S$ with $m$ to form the finite $J$ initial state. Then we need to square and sum over spin. We now apply this to a number
 of decay channels with open charm.

\subsection{$P_c\rightarrow \Lambda_c+\bar D$ decay}

The pentaquark decay through $\Lambda_c$  is larger than through  $\Sigma_c$ or $ \Sigma_c^*$, given the larger access to phase space.
A quick inspection of quantum numbers show that the decay process

$$\bigg[P_c(4440)\bigg[\frac 12\frac 12^-\bigg]_{0}\bigg]\rightarrow \bigg[\Lambda_c(2286)0\frac 12^+\bigg]+\bigg[\bar D(1870)\frac 12 0^-\bigg]$$

\noindent is quadrupolar with $l=2$, since the $l=0$ is forbidden by momentum conservation and $l=1$ by parity. The final meson decay
momentum $|\vec p|\approx 778$ MeV, so the decay produce are non-relativistic.

\subsubsection{$S=0$}

For $S=0$, we need

\begin{align}
{\cal M}(P_c,S=0 \rightarrow D(\epsilon)+\Lambda_c(s))=\alpha \epsilon^{\dagger}\langle 0|a_4+i\vec{a}\cdot \vec{\tau}|\beta \dot \beta\rangle \frac{1}{\sqrt{2}}\sigma_2 \lambda_s \ .
\end{align}
After summing over spin we only need to consider

\begin{align}
\frac{1}{2}\sum_{\beta,\dot \beta} \alpha^2 {\rm tr}\langle 00|a_4+i\vec{a}\cdot \vec{\tau}|\beta \dot \beta\rangle \langle \beta \dot \beta|a_4-i\vec{a}\cdot \vec{\tau}|00\rangle   \ .
\end{align}
The hyper-spherical harmonics $|\beta \dot \beta\rangle$ can be represented in terms of $2\times 2$ matrices as

\begin{align}
\Psi^{l=1}_{\alpha \dot \alpha}(a)=\frac{\sqrt{2}}{\sqrt{\Omega_4}}(\sigma \cdot a)_{\alpha \dot \alpha},
\end{align}
where $\Omega_4=2\pi^2$. We now observe that

\begin{align}
\langle 0|(\bar \sigma \cdot a)_{\dot \alpha,\alpha}|(\sigma \cdot a)^{\beta \dot \beta} \rangle=\frac{\sqrt{2}}{2} \frac{1}{\sqrt{2}}\delta^{\beta}_{\alpha}\frac{1}{\sqrt{2}}\delta^{\dot \beta}_{\dot \alpha} \ .
\end{align}
In terms of these, one has

\begin{align}
\sum_{\beta \dot \beta} \alpha^2 {\rm tr}\langle 00|a_4+i\vec{a}\cdot \vec{\tau}|\beta \dot \beta \rangle \langle \beta \dot \beta|a_4-i\vec{a}\cdot \vec{\tau}|00\rangle =\frac{1}{2}\alpha^2 ,
\end{align}
so that

\begin{align}
\label{DEC1}
\Gamma_{P_c,S=0 \rightarrow D(\epsilon)+\Lambda_c(s)}=\frac{|\vec{p}|}{2\pi} \frac{m_H^{\frac{1}{2}}M_{KK}^{\frac{1}{2}}}{2\tilde\kappa}\frac{1}{4}\langle\alpha\rangle^2 \times \frac{1}{4}\equiv \Gamma
\end{align}
where $\frac{1}{4}$ comes from the initial state averaging over   spin and isospin, and

\begin{align}
\langle \alpha \rangle_{l=1\rightarrow l=0}=4.08/\pi \ .
\end{align}
The decay width (\ref{DEC1})
is fixed by kinematics $|\vec{p}|\sim 778$ MeV, and  the three holographic parameters $\tilde\kappa, M_{KK}, m_H$.

In general, the brane tension for the heavy-light fields $\tilde\kappa$ and that of the light-light fields $\kappa$ are the
same. Here we choose to treat them separatly. In~\cite{Liu:2021tpq} the three parameters $\kappa, M_{KK}, m_H$
were fixed to reproduce globally the charm and bottom baryons as well as the pentaquarks with hidden charm and bottom.
Here, their adjustment to the three observed masses will be subsumed.
The additional $\tilde \kappa$ parameter will be fixed by one measured width as we detail below. All other partial and total widths
will follow in units of $\Gamma$ as predictions, for both charm and bottom.


\subsubsection{$S=1$}

For $S=1$ we need

\begin{align}
&\sum_{S,m,m'}C^{J;\beta+S}_{\beta,S}(C^{J;\beta'+S}_{\beta',S})^{\dagger} \langle\alpha\rangle ^2 {\rm tr}\langle 00|a_4+i\vec{a}\cdot \vec{\tau}|\beta \dot \beta\rangle C^{1S}C^{1S\dagger} \langle \dot \beta \beta'|a_4-i\vec{a}\cdot   \vec{\tau}|00\rangle  \nonumber \\
&=\sum_{s_1,s_2}|\langle S,s_1+s_2|\frac{1}{2},s_1;\frac{1}{2},s_2 \rangle|^2|\langle J,s_1|\frac{1}{2},-s_2;S,s_1+s_2 \rangle|^2  \ .
\end{align}
where we also use $\langle 1,s_1+s_2|\frac{1}{2},s_1;\frac{1}{2},s_2 \rangle$ to denote the expansion coefficient of two spin $\frac{1}{2}$ to one spin 1.
In sum, the decay rate in which all the spins are summed over are the same, and is given by

\begin{align}
&\Gamma_{P_c(J,S)\rightarrow D(\epsilon)+\Lambda_c(s)}=\frac{|\vec{p}|}{2\pi}
\frac{m_H^{\frac{1}{2}}M_{KK}^{\frac{1}{2}}}{2\tilde\kappa}
\frac{1}{4}\langle\alpha\rangle^2 \frac{1}{2(2J+1)}\nonumber  \\
&\times \sum_{s_1,s_2}|\langle S,s_1+s_2|\frac{1}{2},s_1;\frac{1}{2},s_2 \rangle|^2|\langle J,s_1|\frac{1}{2},-s_2;S,s_1+s_2 \rangle|^2 \ .
\end{align}
We define

\begin{align}
f(J,S)=\sum_{s_1,s_2}|\langle S,s_1+s_2|\frac{1}{2},s_1;\frac{1}{2},s_2 \rangle|^2|\langle J,s_1|\frac{1}{2},-s_2;S,s_1+s_2 \rangle|^2   \ ,
\end{align}
with $f(\frac{1}{2},0)=1$ for $S=0$. For $S=1$ we have

\begin{align}
f(J=S\pm \frac{1}{2},S=1)=\sum_{s_1,s_2=\pm \frac{1}{2}}\frac{1}{4}(1+|s_1+s_2|)(1\mp \frac{4}{3}s_1s_2)=\frac{3}{2}\mp \frac{1}{6}
\end{align}
hence  the model independent ratios

\begin{align}
\frac{\Gamma_{P_c(J,S)\rightarrow D(\epsilon)+\Lambda_c(s)}}{\Gamma_{P_c(J',S')\rightarrow D(\epsilon)+\Lambda_c(s)}}=\frac{(2J'+1)f(J,S)}{(2J+1)f(J',S')} \ .
\end{align}
or more explicitly

\begin{align}
\label{RATIOS}
\Gamma\bigg(S=0,J=\frac{1}{2}\bigg):\Gamma\bigg(S=1,J=\frac{1}{2}\bigg):\Gamma\bigg(S=1,J=\frac{3}{2}\bigg)=\frac{1}{2}:\frac{5}{6}:\frac{1}{3}
\end{align}
whenever the decay mode is allowed kinematically.

\subsection{$P_c\rightarrow \Lambda_c+{\bar D}^*$ decay}

$$\bigg[P_c(4440)\bigg[\frac 12\frac 12^-\bigg]_0\bigg]\rightarrow \bigg[\Lambda_c(2286)0\frac 12^+\bigg]+\bigg[{\bar D}^*(2010)\frac 12 1^-\bigg]$$
This decay width can be deduced from that of the scalar meson from the requirement of heavy-quark symmetry. Indeed, the minimal Lagrangian reads in the case of $J=1/2$

\begin{align}
\bar \psi_{P_c} (D+\gamma^5\gamma^{\mu}D_{\mu})\frac{1+\gamma\cdot v}{2}\psi_{\Lambda_c}+c.c \ .
\end{align}
From these the ratio of the decay width for the scalar and the vector are proportional to

\begin{align}
{\rm tr} 1: {\rm tr} \sigma_i\sigma_j \epsilon_i\epsilon_j=1:3 \ .
\end{align}
so that

\begin{align}
\frac{\Gamma_{P_c\rightarrow \Lambda_c+{\bar D}^*}}{\Gamma_{P_c\rightarrow \Lambda_c+\bar D}}=3 \ .
\end{align}

\subsection{$P_c\rightarrow \Sigma_c+\bar D$ decay}

The spin-parity assignment of $\Sigma_c$ is that of $\Lambda_c$ so this decay mode is similar to the one we
addressed earlier

$$\bigg[P_c(4440)\bigg[\frac 12\frac 12^-\bigg]_0\bigg]\rightarrow \bigg[\Sigma_c(2453)1\frac 12^+\bigg]+\bigg[\bar D(1870)\frac 12 0^-\bigg]$$

\noindent which is also quadrupolar with $l=2$. However, the width is expected to be smaller due to the narrower phase space.
The final momentum  is $|\vec p|\approx 502$ MeV,  so again the final kinematics is non-relativistic. To carry the rate,  we need the amplitude

\begin{align}
{\cal M}(P_c,S=0 \rightarrow D(\epsilon)+\Sigma_c(S))=\alpha \epsilon^{\dagger}\langle 1m_1;1m_2|a_4+i\vec{a}\cdot \vec{\tau}|\beta \dot\beta\rangle  \sigma_2\lambda_s C^{\frac{1}{2}S}_{m_1s}\ .
\end{align}
with the corresponding  squared sum

\begin{align}
\sum_{\beta,\dot \beta,m_1,m_1',m_2,S} {\rm Tr}\left[\sigma_2\langle \beta \dot \beta|a_4-i\vec{a}\cdot \vec{\tau}|1m_1;1m_2\rangle (C^{\frac{1}{2}S}_{m_1})^{\dagger}C^{\frac{1}{2}S}_{m_1'}\langle 1m_1';1m_2  |a_4+i\vec{a}\cdot \vec{\tau}|\beta \dot \beta\rangle\sigma_2\right] \ .
\end{align}
For the $S=1$ state  we need the  re-summation

\begin{align}
&\sum_{\beta,\beta',\dot \beta, m_1,m_1',m_2,S,S'}C^{J;\beta+S}_{\beta,S}(C^{J;\beta'+S}_{\beta,S})^{\dagger}\nonumber \\ &\times{\rm Tr}\left[C^{1S\dagger}\langle \beta \dot\beta|a_4-i\vec{a}\cdot \vec{\tau}|1m_1;1m_2\rangle (C^{\frac{1}{2}S'}_{m_1})^{\dagger}C^{\frac{1}{2}S'}_{m_1'}\langle 1m_1'; 1m_2|a_4+i\vec{a}\cdot \vec{\tau}|\beta'\dot \beta\rangle C^{1S}\right]\ .
\end{align}
This can be achieved using the following identity in terms of  Clebsch-Gordon coefficients

\begin{align}
\langle 1,m_1;1,m_2| (\bar \sigma \cdot a)_{\dot \alpha \alpha}|(\sigma \cdot a)^{\beta \dot \beta}\rangle=\mathbb A \langle 1,m_1|\frac{1}{2},\alpha;\frac{1}{2},-\beta\rangle\langle 1,m_2|\frac{1}{2},\dot \alpha;\frac{1}{2},-\dot \beta\rangle \ ,
\end{align}
where $\mathbb A$ is a numerical number independent of spin. The minus sign follows from lowering $\beta$ and $\dot \beta$
down using $\sigma_2$ to form the spin-sum.  To  evaluate  $\mathbb A$, we may  choose $\alpha=\dot \alpha=+$, and $\beta=\dot \beta=-$, then sum over $m_1$ and $m_2$, to obtain

\begin{align}
|\mathbb A|^2=\frac{2}{\Omega_4}\sum_{P_2}\bigg(\int d\Omega_4 (a_1^2+a_2^2)\Phi_{P_2}(a)\bigg)^2 =\frac{1}{6}\ ,
\end{align}
where the sum over $P_2$ ranges over all the $9$ independent hyper-spherical harmonic functions for $l=2$.

Using the above results, the decay rate can be obtained by summing over all the spins,

\bea
&&\Gamma_{P_c(J,S)\rightarrow \Sigma_c+D}=\frac{|\vec{p}|m_H^{\frac{1}{2}}M_{KK}^{\frac{1}{2}}}{4\pi \tilde \kappa} \frac{\langle \alpha \rangle^2}{6(2J+1)}|\mathbb A|^2\nonumber\\
&&\times \sum_{s_1,s_2}|\langle S,s_1+s_2|\frac{1}{2},s_1;\frac{1}{2},s_2 \rangle|^2|\langle J,s_1|\frac{1}{2},-s_2;S,s_1+s_2 \rangle|^2\nonumber\\
&&\qquad\times|\langle 1,0|\frac{1}{2},s_1;\frac{1}{2},-s_1\rangle|^2|\langle \frac{1}{2},-s_1|1,0;\frac{1}{2},-s_1\rangle|^2 \nonumber \\
&&=\frac{|\vec{p}|m_H^{\frac{1}{2}}M_{KK}^{\frac{1}{2}}}{4\pi \tilde\kappa}\frac{\langle \alpha \rangle^2}{36(2J+1)}|\mathbb A|^2\nonumber\\
&&\times \sum_{s_1,s_2}|\langle S,s_1+s_2|\frac{1}{2},s_1;\frac{1}{2},s_2 \rangle|^2|\langle J,s_1|\frac{1}{2},-s_2;S,s_1+s_2 \rangle|^2  \ ,
\eea
where the  additional factor of  $\frac{1}{6}$ originates  from

\begin{align}
|\langle 1,0|\frac{1}{2},s_1;\frac{1}{2},-s_1\rangle|=\frac{1}{\sqrt{2}} \ , \qquad\qquad
|\langle \frac{1}{2},-s_1|1,0;\frac{1}{2},-s_1\rangle|=\sqrt{\frac{2}{3}}\times \frac{1}{\sqrt{2}} \ .
\end{align}
This decay rate relates to the one for $\Lambda_c$, and the model independent ratio is

\begin{align}
\frac{\Gamma_{P_c(J,S)\rightarrow \Sigma_c+D}}{\Gamma_{P_c(J,S)\rightarrow \Lambda_c+D}}=\frac{502}{778}\times \frac{8}{36}\times \frac{4.97^2}{4.08^2}=0.574 \ .
\end{align}

\subsection{$P_c\rightarrow \Sigma^*_c+\bar D$ decay}

$$\bigg[P_c(4440)\bigg[\frac 12\frac 12^-\bigg]_0\bigg]\rightarrow \bigg[\Sigma^*_c(--)1\frac 12^-\bigg]+\bigg[\bar D(1870)\frac 12 0^-\bigg]$$

\noindent with $l=1$ by parity. In this case the formula remains the same as the preceding one,
with the only change being the value of $n_z$ in the averaging  over $\alpha$

\bea
&&\Gamma_{P_c(J,S)\rightarrow \Sigma^{\star}_c+D}=\frac{|\vec{p}|}{8\pi M_{P_c}^2} \frac{\langle \alpha \rangle^2}{(2J+1)}|\mathbb A|^2\nonumber\\
&&\times \sum_{s_1,s_2}|\langle S,s_1+s_2|\frac{1}{2},s_1;\frac{1}{2},s_2 \rangle|^2|\langle J,s_1|\frac{1}{2},-s_2;S,s_1+s_2 \rangle|^2\nonumber\\
&&\qquad\times|\langle 1,0|\frac{1}{2},s_1;\frac{1}{2},-s_1\rangle|^2|\langle \frac{1}{2},-s_1|1,0;\frac{1}{2},-s_1\rangle|^2 \nonumber \\
&&=\frac{|\vec{p}|}{8\pi M^2} \frac{\langle\alpha\rangle^2}{6(2J+1)}|\mathbb A|^2\nonumber\\
&&\times \sum_{s_1,s_2}|\langle S,s_1+s_2|\frac{1}{2},s_1;\frac{1}{2},s_2 \rangle|^2|\langle J,s_1|\frac{1}{2},-s_2;S,s_1+s_2 \rangle|^2  \ ,
\eea
However, since the in-coming and out-going states have  different parity in the z-direction, the average of $\alpha$ will be zero in this case, hence

\begin{align}
\Gamma_{P_c(J,S)\rightarrow \Sigma^{\star}_c+D}=0 \ .
\end{align}

\section{Hidden charm decay}~\label{HIDDEN}

The $P_c(4440)$ state can strongly decay only through  $J/\Psi$ with hidden charm because of kinematics,

$$\bigg[P_c(4440)\bigg[\frac 12\frac 12^-\bigg]_0\bigg]\rightarrow \bigg[J/\Psi(3097)01^-\bigg]+\bigg[p(938)\frac 12\frac 12^+\bigg]$$

\noindent with $l=0,2$. The decay momentum for charm is about $|\vec{P}|\approx 809$ MeV, so the final kinematics is relativistic. To determine
the transition coupling  $P_c\rightarrow J/\Psi+p$ we needs the U(1)  transition current

\be
\label{TFF}
\left<P_c\bigg[p_2, \frac 12 \frac 12^-\bigg]\bigg|{\mathbb J}^\mu(0)\bigg|P\bigg[p_1, \frac 12 \frac 12^+\bigg]\right> \ ,
\ee
in which the in-out states in (\ref{TFF}) are eigenstates of the moduli Hamiltonian defined earlier.

\subsubsection{Bulk-to-boundary current}

To determine (\ref{TFF}), we consider the decay of pentaquarks into $J/\Psi$ (Upsilon)  represented by U(1)  vector field $\delta A_\mu(z)e^{-2im_Ht}$ and the nucleon. To obtain the change in the Lagrangian one needs to select the terms that mixes the quark and anti-quarks. One first consider $\delta A_0$, this leads to the temporal coupling

\begin{align}
\delta L_T=\delta A_0\frac{1}{m_H}\tilde \rho_1 \ ,
\end{align}
where

\begin{align}
\tilde \rho_1=\left(-\frac{9}{4\tilde \rho^2}f^2+\frac{3}{16\pi^3 a}\frac{2\rho^2-X^2}{(X^2+\rho^2)^2}f^2\right)\overline u_Q v_{\bar Q}\chi_Q^\dagger\chi^\dagger_{\bar Q}+{\rm h.c.}
\end{align}
On the other-hand, one also needs to consider $\delta A_M$ which contributes actually at leading order in $\lambda$, but next to leading order in ${1}/{m_H}$. This
amounts to a spatial coupling through

\begin{align}
\delta L_{S}=4aN_c\lambda \delta A_M \partial _N\left(\Phi^{\dagger}_M \Phi_N-\Phi^{\dagger}_N \Phi_M\right) \ ,
\end{align}
which is of order  ${1}/{m_H}$  in the heavy quark limit.
$A_M$ sources a U(1) gauge field with bulk vector modes satisfying~\cite{Sakai:2004cn}

\begin{align}
\label{VECTORn}
-(1+Z^2)^{\frac{1}{3}}\partial_Z((1+Z^2)\partial_Z \varphi_n(Z))=\lambda_n \varphi_n
\end{align}
and  normalized according to

\begin{align}
\int dZ \frac{1}{(1+Z^2)^{\frac{1}{3}}}|\varphi_n(Z)|^2=1 \ ,
\end{align}
Recall that  in the Sakai-Sugimoto model, the holographic coordinate
$Z={z}/{U_{KK}}$  with $U_{KK}\propto M_{KK}$. In the light-light sector, $M_{KK}$ is fixed to reproduce the
low-lying rho meson states $\tilde m_n=\lambda_nM_{KK}$ (odd $n$)~\cite{Sakai:2004cn}.

In terms of the eigen-modes (\ref{VECTORn}), the bulk-to-bulk vector propagator is given by

\begin{align}
G_{MN}(E;\vec{P},Z,X;Z',X')=\frac{g_{MN}}{\tilde \kappa}\sum_{n}\frac{\varphi_n(Z)\varphi^{\dagger}_n(Z')e^{-i\vec{P}\cdot(\vec{X}-\vec{X}')}}{E^2-\vec{P}^2-\tilde m_n^2} \ ,
\end{align}
and the  bulk-to-boundary U(1) gauge field is

\begin{align}
A^{M}(E;Z,\vec x-\vec X)=\frac{e^{-i\vec{P}\cdot(\vec x-\vec X)}}{\sqrt{\tilde \kappa}}\sum_{n}\frac{\varphi_n(Z)a_n^M}{E^2-\vec{P}^2-\tilde m_n^2} \ ,
\end{align}
for the spatial components $M=1,2,3,z$ with the bulk modular sources

\bea
a_n^Z&=&\frac{4\kappa}{8m_HaN_c}i\vec{P}\cdot \overline{u}_Q\vec\sigma v_{\bar Q}\,\chi^\dagger_Q\chi_{\bar Q}^\dagger+{\rm h.c.}\nonumber\\
\vec{a_n}&=&\frac{4\kappa}{8m_HaN_c}i\vec{P}\times \overline{u}_Q\vec\sigma v_{\bar Q}\,\chi^\dagger_Q\chi_{\bar Q}^\dagger+{\rm h.c.}
\eea
with $\chi^\dagger_{Q,\bar Q}$ fermionic creation operators in the pentaquark moduli satisfying  anti-commutation relations~\cite{Liu:2017xzo}.
Here we have used the normalization condition $\int dZdX f^2(Z^2+X^2)=1$.
Similarly, the bulk-to-boundary temporal component $A^0$  with full back reaction is

\begin{align}
A^0(E;Z,\vec x-\vec X)=\frac{e^{-i\vec{P}\cdot(\vec x-\vec X)}}{ \sqrt{\tilde \kappa}}\sum_{n}\frac{\varphi_n(Z)a_n^0}{E^2-\vec{P}^2-\tilde m_n^2} \ ,
\end{align}
with the modular source

\begin{align}
a_n^0=\frac{4\tilde \kappa}{aN_cm_H}\int dZ d^3X \left(-\frac{9}{4\tilde \rho^2}f^2+\frac{3}{16\pi^3 a}\frac{2\rho^2-X^2}{(X^2+\rho^2)^2}f^2\right)
 \overline{u}_Qv_{\bar Q}\,\chi^\dagger_Q\chi_{\bar Q}^\dagger+{\rm h.c.}
\end{align}
The boundary U(1) current sourced by the topological pentaquark in bulk, follows from  the canonical identification~\cite{Sakai:2004cn}

\begin{align}
\vec{\mathbb J}=-\tilde \kappa \vec{F}^z|_{z=-\infty}^{z=\infty} \ ,
\end{align}
which is

\begin{align}
\label{JXX}
\vec{\mathbb J}(\vec{x}-\vec X)=-\sum_{n}\frac{ \lambda g_n \varphi_n(Z)}{4m_H\sqrt{\tilde \kappa}}\int \frac{d^3\vec{P}e^{-i\vec{P}\cdot(\vec x-\vec X)}}{E^2-\vec{P}^2-\tilde m_n^2}
i\vec{P}\times \overline{u}_Q\vec\sigma v_{\bar Q}\,\chi^\dagger_Q\chi_{\bar Q}^\dagger +{\rm h.c.}
\end{align}
The pentaquark U(1) current at the boundary is sourced by the spin of the emerging $Q\bar Q$ attachment in bulk to order $1/m_H$, with a $1^{--}$ vector cloud
composed essentialy of the rho-meson Regge trajectory.  This is not surprising given the holographic spin transmutation $\vec J\rightarrow \vec J+\vec S_Q$ discussed in~\cite{Liu:2017xzo}. This is the first major result in this section.

\subsubsection{Transition amplitude and width}

In terms of the boundary current (\ref{JXX}), the transition form factor $P_c\rightarrow V +p$ reads

\begin{align}
\label{JPP}
\langle P|{\vec {\mathbb J}}(\vec{x}-\vec X)|P_c\rangle=(i\vec{P}\times  \overline{v}_{\bar Q}\vec\sigma u_{Q}) G(\vec{P}) (2\pi)^3\delta^3(P'-P)\ ,
\end{align}
with the induced vector form factor

\begin{align}
\label{GP}
G(\vec{P})=\lambda \sqrt{\frac{m_N}{M_{P_c}}}\sum_n\frac{\langle \varphi_n(Z)\rangle}{\sqrt{\tilde \kappa}}\frac{g_n}{E^2-\vec{P}^2-\tilde m_n^2}
\end{align}
The averaging in (\ref{GP}) is over the Gaussian baryonic (nucleon and pentaquark)  modular wavefunctions which are localized around $Z\sim 0$~\cite{Liu:2017xzo}.

 For comparison, we note that
 in the soft wall model, the bulk-to-boundary propagator  $G(P,Z)$ can be expressed in terms of confluent hypergeometric functions ${\cal U}$ as

\begin{align}\label{eq:formfactorepicit}
G(P,Z)\sim  M_{KK}^2Z^2\lambda\sqrt{\frac{m_N}{M_{P_c}}}\Gamma\bigg(1-\frac{P^2}{4M_{KK}^2}\bigg){\cal U}\bigg(1-\frac{-P^2}{4M_{KK}^2};2;M_{KK}^2Z^2\bigg) \ .
\end{align}
The form factor follows by averaging over the Dirac fields in bulk. In contrast, the latters  are localized around $Z\sim \infty$
to satisfy the hard scattering rules.

The scattering amplitude follows  by  LSZ reduction of (\ref{JPP})

\begin{align}
{\cal M}=\vec{\epsilon}^{\star}\cdot(i\vec{P}\times  \overline{v}_{\bar Q}\vec\sigma u_{Q}) \lambda \sqrt{\frac{m_N}{M_{P_c}}}\frac{\langle\varphi_n(Z) \rangle}{\sqrt{\tilde \kappa}}
\end{align}
for the emitted vector meson labeled by $V=n$.
The squared scattering amplitude after summing over the polarizations, reads

\begin{align}
|{\cal M}|^2=\frac{|\tilde G|^2|\vec{P}|^2}{2S+1}\sum_{s_1,s_2;s_1's_2'}(\vec{n}\times \vec{\sigma})_{s_1s_2}\cdot (\vec{n}\times \vec{\sigma})_{s_2's_1'}  \sum_{M_1}C^{SM_1}_{s_1s_2}C^{SM_1}_{s_1's_2'}
\end{align}
with
\be
\tilde G(\vec{P})= \lambda \sqrt{\frac{m_N}{M_{P_c}}}\frac{\langle\varphi_n(Z) \rangle}{\sqrt{\tilde \kappa}}
\ee
which can be reduced

\begin{align}
|{\cal M}|^2&=|\tilde G|^2|\vec P|^2\sum_{M_1;s_1s_2;s_1's_2'}\frac{\vec{\sigma}_{s_1s_2}\cdot \vec{\sigma}_{s_2's_1'}-\sigma^3_{s_1s_2}\sigma^3_{s_1's_2'}}{2S+1} C^{SM_1}_{s_1s_2}C^{SM_1}_{s_1's_2'} \nonumber \\
&=|\tilde G|^2|\vec P|^2\sum_{M_1;s_1s_2;s_1's_2'}\frac{2\delta_{s_1s_1'}\delta_{s_2s_2'}-\delta_{s_1s_2}\delta_{s_1's_2'}-\sigma^3_{s_1s_2}\sigma^3_{s_1's_2'}}{2S+1}C^{SM_1}_{s_1s_2}C^{SM_1}_{s_1's_2'} \nonumber\\
&=\frac{2|\vec{P}|^2}{3} |\tilde G|^2 \,(\delta_{S=1}+3\delta_{S=0})\ .
\end{align}
Since the heavy quark in the initial state is still non-relativistic, the decay rate for $P_c\rightarrow \gamma p$ is then

\begin{align}
\Gamma=|\vec{P}|\times \frac{ |\vec{P}|^2}{4\pi M_{P_c}^2} \frac{|\tilde G|^2}{2S+1} \ ,
\end{align}
or
\begin{align}
\Gamma=|\vec{P}|\times \frac{ |\vec{P}|^2}{4\pi(2S+1) M_{P_c}^2} \frac{\lambda^2 m_N}{M_{P_c}} \times \bigg|\frac{\langle \varphi_n(Z)\rangle}{\sqrt{\tilde \kappa}}\bigg|^2 \ .
\end{align}
The $J/\Psi$ bulk wavefunction satisfies a vector equation similar to (\ref{VECTORn}) except for the overall scale. Indeed,
recall that  in the Sakai-Sugimoto construction $Z={z}/{U_{KK}}$ and $U_{KK}\propto M_{KK}$,
which is usually fixed by the light vector meson rho mass. For $J/\Psi$ we  set
$M_{KK}\rightarrow 2m_H$ to the heavy meson mass. As a result, the bulk $J/\Psi$ wavefunctions follow from the bulk rho wavefunctions by rescaling

\begin{align}
\varphi_n(Z)\rightarrow \sqrt{\frac{M_{KK}}{2m_H}}\varphi_n\bigg(\frac{M_{KK}}{2m_H} Z\bigg),
\end{align}
which leads to the partial decay width

\begin{align}
\label{GAMHEAVY}
\Gamma=|\vec{P}|\times \frac{ |\vec{P}|^2}{4\pi(2S+1) M_{P_c}^2} \frac{\lambda^2 m_N}{M_{P_c}} \times \frac{M_{KK}}{M_{P_c}}
\bigg(\frac{\varphi_n(0)}{\sqrt{\tilde\kappa}}\bigg)^2\ .
\end{align}
in the heavy quark limit.

We note the further suppression by $1/m_H$ of the hidden decay width (\ref{GAMHEAVY}) in comparison to the open decay
widths derived earlier. Indeed, a  comparison  with the open channel decay width  yields the ratio

\begin{align}
\label{S1X}
\frac{\Gamma_{P_c\rightarrow J/\psi+P}}{\Gamma_{P_c\rightarrow \Lambda_c+\bar D}}=\lambda^2\bigg(\frac{16\sqrt{2}}{2S+1}\bigg)
\bigg(\frac{|\vec{P}|^3m_NM_{KK}^{\frac{3}{2}}}{|\vec{p}|M_{P_c}^\frac{9}{2}}\bigg) \bigg(\frac{|\varphi_n(0)|^2}{\langle \alpha\rangle ^2}\bigg) \ .
\end{align}
The mismatch in the kinematical momenta in the ratio reflects on the fact that the pentaquark decay to hidden charm
follows from a Pauli-like coupling, while all open charm decays proceed from a Dirac-like coupling. For $P_c(440)$,
the decay kinematics fixes  $|\vec{P}|\approx 809$ MeV and $|\vec{p}|\approx 778$ MeV. Using $M_{KK}=0.495$ MeV
and  $\lambda=g_{YM}^2N_c=10$, the ratio  (\ref{S1X})  is

\begin{align}
\frac{\Gamma_{P_c\rightarrow J/\psi+P}}{\Gamma_{P_c\rightarrow \Lambda_c+\bar D}}=\frac{0.34|\varphi_n(0)|^2}{2S+1} \ .
\end{align}
The numerical value of the vector wave function at the origin solution to (\ref{VECTORn}),
is about $\frac{1}{2}$ for the ground state with $n=1$, so that

\begin{align}
\frac{\Gamma_{P_c\rightarrow J/\psi+P}}{\Gamma_{P_c\rightarrow \Lambda_c+\bar D}}\sim \frac{0.085}{2S+1} \ ,
\end{align}
The  decay width in the hidden channel is about $\frac{1}{10}$ the one observed in the open channels. This observation
is in qualitative agreement with the one made using molecular bound states~\cite{Eides:2018lqg,Lin:2019qiv}.

For completeness and clarity, we have  collected  all the partial decay widths for charm  pentaquark states,
including their total witdth in units of $\Gamma$ (\ref{DEC1}) in Tables~\ref{tab_4440}-\ref{tab_4457}-\ref{tab_4312}.
Overall, the decay widths of $P_c(4440)$ and $P_c(4312)$ are found to be comparable, while the total decay width
of $P_c(4457)$ is smaller. Within error bars, these observations are compatible with the charm pentaquark widths
(\ref{LHCBMW}) reported by LHCb. To fix the value of $\Gamma$ in (\ref{DEC1}) (equivalently the value of the holographic parameter $\tilde\kappa$) and therefore all the remaining widths listed in the tables,
we use the measured central value of the  total width
of $P_c(4440)$ in (\ref{LHCBMW}), namely
$$\Gamma=\frac{20.6\pm 4.9\,{\rm MeV}}{4.66}=4\pm 1 {\rm MeV}$$

The yet to be observed bottom pentaquarks and their widths are listed  in Table~\ref{tab_widthb}.
For the bottom results, we used
$m_{H}=5111$ MeV  fixed by
the heavy-light B-meson mass in (\ref{MDMH}).
For the bottom Pentaquark mass, we use   the central holographic value $M_{P_b}=11163$ MeV~\cite{Liu:2021tpq},
as the predicted three holographic bottom pentaquark masses are very close in mass. The differences in the widths
listed stem from the different spin assignments.
The  broader  width for bottom versus charm recorded in the Tables
$${[P_b\rightarrow \Lambda_b\bar B]\over [P_c\rightarrow \Lambda_c\bar D]}\sim 2.58$$
stems from the larger momentum of the decay produce and the larger value for $m_H$.   The much smaller ratio
$${[P_b\rightarrow \Upsilon p]\over [P_c\rightarrow J/\Psi p]}=\bigg(\frac{M_{P_c}}{M_{P_b}}\bigg)^{\frac 92}\sim 0.02$$
for fixed momentum decay,  follows from  the larger suppression by the bottom Pentaquark mass.

\begin{widetext}
\begin{table}[h]
\caption{Pentaquark $P_c(4440)[\frac 12\frac12^-]_0$ decay widths in units of $\Gamma$}
\begin{center}
\begin{tabular}{ccccccccc}
\hline
\hline
Decay mode  & Final momentum (MeV) &  Width  \\
\hline
\hline
$P_c\rightarrow \Lambda_c\bar D$    & 778 MeV  & 1& \\
$P_c\rightarrow \Sigma_c\bar D$     & 502 MeV   & 0.574& \\
$P_c\rightarrow \Lambda_c{\bar D}^*$   & 778 MeV &3 & \\
$P_c\rightarrow \Sigma^*_c\bar D$     & --- MeV  & 0& \\
\hline
$P_c\rightarrow J/\Psi p$     & 809 MeV   & 0.085& \\
\hline
Total width & & 4.66&\\
\hline
\hline
\end{tabular}
\end{center}
\label{tab_4440}
\end{table}%
\end{widetext}

\begin{widetext}
\begin{table}[h]
\caption{Pentaquark $P_c(4457)[\frac 12\frac32^-]_1$ decay widths in units of $\Gamma$}
\begin{center}
\begin{tabular}{ccccccccc}
\hline
\hline
Decay mode  & Final momentum (MeV) &  Width  \\
\hline
\hline
$P_c\rightarrow \Lambda_c\bar D$    & 801 MeV  & 0.68& \\
$P_c\rightarrow \Sigma_c\bar D$     & 537 MeV   & 0.409& \\
$P_c\rightarrow \Lambda_c{\bar D}^*$   & 801 MeV &2.04 & \\
$P_c\rightarrow \Sigma^*_c\bar D$     & --- MeV  & 0& \\
\hline
$P_c\rightarrow J/\Psi p$     & 828 MeV   & 0.043& \\
\hline
Total width & & 3.172&\\
\hline
\hline
\end{tabular}
\end{center}
\label{tab_4457}
\end{table}%
\end{widetext}

\begin{widetext}
\begin{table}[h]
\caption{Pentaquark $P_c(4312)[\frac 12\frac12^-]_1$ decay widths in units of $\Gamma$}
\begin{center}
\begin{tabular}{ccccccccc}
\hline
\hline
Decay mode  & Final momentum (MeV) &  Width  \\
\hline
\hline
$P_c\rightarrow \Lambda_c\bar D$    & 571 MeV  & 1.22 & \\
$P_c\rightarrow \Sigma_c\bar D$     & --- MeV   & 0& \\
$P_c\rightarrow \Lambda_c{\bar D}^*$   & 571 MeV &3.66& \\
$P_c\rightarrow \Sigma^*_c\bar D$     & --- MeV  & 0& \\
\hline
$P_c\rightarrow J/\Psi p$     & 658MeV   & 0.014& \\
\hline
Total width & & 4.894&\\
\hline
\hline
\end{tabular}
\end{center}
\label{tab_4312}
\end{table}%
\end{widetext}

\begin{widetext}
\begin{table}[h]
\caption{Pentaquark $P_b(11163)[[\frac 12\frac12^-]_0/[\frac 12\frac12^-]_1/[\frac 12\frac32^-]_1]$ decay widths in units of $\Gamma$}
\begin{center}
\begin{tabular}{ccccccccc}
\hline
\hline
Decay mode & Final momentum (MeV) &  Width  \\
\hline
\hline
$P_b\rightarrow \Lambda_b\bar B$     & 1206 MeV  & 2.38/3.96/1.58& \\
$P_b\rightarrow \Sigma_b\bar B$     & 640 MeV   &1.21/2.01/1.81 \\
$P_b\rightarrow \Lambda_b{\bar B}^*$     & 1260 MeV & 7.14/11.9/4.76& \\
$P_b\rightarrow \Sigma^*_b\bar B$     & --- MeV  &0& \\
\hline
$P_b\rightarrow \Upsilon p$     & 1310MeV   & 0.006/0.002/0.002& \\
\hline
Total width & & 10.76/17.87/8.15&\\
\hline
\hline
\end{tabular}
\end{center}
\label{tab_widthb}
\end{table}%
\end{widetext}


\section{Conclusions}~\label{CONCLUSION}

In leading order in the heavy quark mass $m_H$, the holographic  construction predicts three heavy pentaquark states with the
assignments $[\frac 12\frac 12^-]_{S=0,1}$  and $[\frac 12\frac 32^-]_{S=1}$~\cite{Liu:2017xzo,Liu:2017frj}, which are BPS,  degenerate and stable by
heavy quark symmetry. In this limit, the heavy-light  $[0^-,1^-]=[D, D^*]$ multiplet  binds democratically to an instanton core in bulk  with equal
spin and isospin. The core is stable by dual  gauge-gravity interactions, and the ensuing dynamics has manifest chiral and heavy quark symmetries.
The construction has very few parameters (three) with no need for ad-hoc form factors.

The existence of three instead of two pentaquark states as originally reported,
is compatible with the recent  re-analysis by the LHCb collaboration~\cite{Aaij:2019vzc}, although the quantum number assignments are yet
to be  identified experimentally.
The newly reported $P_c(4337)$~\cite{LHCb:2021chn} state appears too low and narrow  for an  excited holographic pentaquark state $P^*$ candidate~\cite{Liu:2017frj,Liu:2021tpq}.  We also
expect the chiral pentaquark doublers following from the addition of
the mirror multiplet $[0^+,1^+]=[\tilde D, \tilde D^*]$~\cite{Nowak:1992um,Bardeen:2003kt,Liu:2016kqx},
to be more massive and  even unbound.


We have shown how to systematically organize the spin corrections  using the holographic bound state approach to the
pentaquark states, away from the heavy quark mass  limit.  To order  $1/m_H$, spin effects lift the mass degeneracy through spin-orbit effects~\cite{Liu:2021tpq},
and the pentaquark states undergo strong decays in channels with open and hidden charm (bottom).
We have explicitly derived the spin induced vertices and  used them to construct the pertinent  transition amplitutes and form factors. Some
of the transition form factors, e.g. $\gamma +p\rightarrow P_c$ may be accessible to precision photo- or electro-excitations of
Pentaquarks~\cite{Wang:2015jsa,Kubarovsky:2015aaa,Karliner:2015voa} as currently pursued at JLab~\cite{Meziani:2020oks}.

The transition couplings and form factors drive the strong decay widths of both charm
and bottom pentaquarks, which are  tied by symmetry to a single decay mode say $P_c\rightarrow \Lambda_c+\bar D$.
In particular,  the partial widths of the three pentaquark  states are found to satisfy  model independent ratios,
whenever allowed by kinematics.
These observations carry to the bottom pentaquark states as well.
The  holographic analysis of the pentaquark states with hidden charm and bottom  is  extemely predictive and thus falsifiable.


\vskip 1cm
{\bf Acknowledgements}

This work is supported by the Office of Science, U.S. Department of Energy under Contract No. DE-FG-88ER40388,
and by the Polish National Science Centre (NCN) Grant UMO- 2017/27/B/ST2/01139.

\appendix

\section{Moduli coefficient $\alpha$}~\label{COEFFICIENT}

The moduli coefficient entering the Hamiltonian (\ref{HAMX}) follows  from the averaging over the collective coordinates
of the instanton

\begin{align}
\label{COEF1}
\alpha\bigg(\rho,\frac{Z}{\rho}\bigg)&=\frac{N_c}{2\pi^2\sqrt{16aN_c}}\int d\tilde Z d^3\tilde X\frac{c\phi_n(\tilde Z+Z)}{(\tilde X^2+\tilde Z^2+\rho^2)^{\frac{5}{2}}}\left(1+\frac{5\rho^2}{2(\tilde X^2+\tilde Z^2+\rho^2)}\right) \nonumber \\
&\rightarrow \frac{\rho N_c}{\sqrt{2}\pi^3\sqrt{16aN_c}}\int dZ \phi_n(Z)\int d^3\tilde X \frac{1}{(\tilde X^2+Z^2+\rho^2)^{\frac{5}{2}}}\left(1+\frac{5\rho^2}{2(\tilde X^2+Z^2+\rho^2)}\right) \nonumber \\
&=\frac{\sqrt{2}\rho N_c}{3\pi^2\sqrt{aN_c}(Z^2+\rho^2)}\int dZ \phi_n(Z) \ ,
\end{align}
and depends both on the instanton size $\rho$ and the holographic $Z$-coordinate on the moduli. For our case we only need the modular wavefunction
$n=1$, for which

\begin{align}
\int dZ \phi_0(Z)=2^{\frac{5}{8}}\pi^{\frac{1}{4}} \frac{1}{\sqrt{2\kappa}m_H^{\frac{1}{4}}m_{KK}^{\frac{1}{4}}}\ ,
\end{align}
and

\begin{align}
\bigg<\frac{\sqrt{2}\rho N_c}{3\pi^2\sqrt{aN_c}(Z^2+\rho^2)}\bigg>\sim \frac{4\sqrt{2}N_c}{3\pi}\bigg<\frac{\tilde \rho}{\tilde \rho^2+\tilde Z^2} \bigg> \ .
\end{align}
To carry the the $\rho$-expectation value we  need  the radial wave functions for $l=1$, $l=0$, $l=2$

\begin{align}
R_{l=1,0,2}(\tilde \rho)=\tilde \rho^{-1+\sqrt{(l+1)^2+\frac{36}{5}}}e^{-\frac{\tilde \rho^2}{\sqrt{6}}} \ .
\end{align}
and  the  modular wavefunction (non-normalized)

\begin{align}
\psi(Z)=e^{-\frac{Z^2}{\sqrt{6}}} \ .
\end{align}
The results for  $k=0$  are

\begin{align}
\langle l=0|\frac{\tilde \rho}{\tilde \rho^2+\tilde Z^2} |l=1\rangle=0.35 \ , \\
\langle l=0|\frac{\tilde \rho}{\tilde \rho^2+\tilde Z^2} |l=1\rangle=0.43 \ .
\end{align}
The ensuing numerical values associated to the transition coefficients in (\ref{HAMX}) are

\begin{align}
\langle \alpha \rangle_{l=1\rightarrow l=0}=4\sqrt{2}\times 0.35 \times 2^{\frac{5}{8}}\pi^{\frac{1}{4}}/\pi=4.08/\pi \ , \\
\langle \alpha \rangle_{l=1\rightarrow l=2}=4\sqrt{2}\times 0.43 \times 2^{\frac{5}{8}}\pi^{\frac{1}{4}}/\pi=4.97/\pi \ .
\end{align}

\section{Properties of Clebsch-Gordon Coefficients }

Here we detail our conventions for the Clebsch-Gordon coefficients used. We denote  by  $|j_1m_1\rangle$ and $|j_2m_2\rangle$ the state-vector for the standard $2j_1+1$ and $2j_2+1$ irreducible representations of the $su(2)$ Lie algebra. The  tensor product splits into $J=|j_1-j_2|,...|j_1+j_2|$ irreducible representations in the following way

\begin{align}
|JM\rangle= \sum_{m_1,m_2}|j_1m_1\rangle |j_2m_2\rangle \langle j_1m_1;j_2m_2|JM\rangle \ ,
\end{align}
where $\langle j_1m_1;j_2m_2|JM\rangle$ are the Clebsch-Gordon coefficients normalized according to

\begin{align}
\sum_{m_1,m_2}|\langle j_1m_1;j_2m_2|JM\rangle|^2=1  \ .
\end{align}
For simplicity we write

\begin{align}
\langle j_1m_1;j_2m_2|JM\rangle \equiv C^{JM}_{m_1m_2} \ .
\end{align}
To carry the sums in the text, we make use of the orthogonality relations, and
the following symmetry properties

\bea
|\langle j_1m_1;j_2m_2|JM\rangle|&=&|\langle j_2m_2;j_1m_1|JM\rangle| \ , \\
|\langle j_1m_1;j_2m_2|JM\rangle|&=&\sqrt{\frac{2J+1}{2j_1+1}}|\langle J(-M);j_2m_2 |j_1(-m_1)\rangle| \ ,
\eea
as well as the explicit  relation

\begin{align}
|\langle j(M-\frac{s}{2});\frac{1}{2}\frac{s}{2}|(j\pm \frac{1}{2})M\rangle|=\sqrt{\frac{1}{2}\bigg(1\pm\frac{sM}{j+\frac{1}{2}}\bigg)} \ ,
\end{align}
if $j_1$ or $j_2$ are equal to $\frac{1}{2}$ .

\bibliography{HL}

\begin{thebibliography}{33}%
\makeatletter
\providecommand \@ifxundefined [1]{%
 \@ifx{#1\undefined}
}%
\providecommand \@ifnum [1]{%
 \ifnum #1\expandafter \@firstoftwo
 \else \expandafter \@secondoftwo
 \fi
}%
\providecommand \@ifx [1]{%
 \ifx #1\expandafter \@firstoftwo
 \else \expandafter \@secondoftwo
 \fi
}%
\providecommand \natexlab [1]{#1}%
\providecommand \enquote  [1]{``#1''}%
\providecommand \bibnamefont  [1]{#1}%
\providecommand \bibfnamefont [1]{#1}%
\providecommand \citenamefont [1]{#1}%
\providecommand \href@noop [0]{\@secondoftwo}%
\providecommand \href [0]{\begingroup \@sanitize@url \@href}%
\providecommand \@href[1]{\@@startlink{#1}\@@href}%
\providecommand \@@href[1]{\endgroup#1\@@endlink}%
\providecommand \@sanitize@url [0]{\catcode `\\12\catcode `\$12\catcode
  `\&12\catcode `\#12\catcode `\^12\catcode `\_12\catcode `\%12\relax}%
\providecommand \@@startlink[1]{}%
\providecommand \@@endlink[0]{}%
\providecommand \url  [0]{\begingroup\@sanitize@url \@url }%
\providecommand \@url [1]{\endgroup\@href {#1}{\urlprefix }}%
\providecommand \urlprefix  [0]{URL }%
\providecommand \Eprint [0]{\href }%
\providecommand \doibase [0]{http://dx.doi.org/}%
\providecommand \selectlanguage [0]{\@gobble}%
\providecommand \bibinfo  [0]{\@secondoftwo}%
\providecommand \bibfield  [0]{\@secondoftwo}%
\providecommand \translation [1]{[#1]}%
\providecommand \BibitemOpen [0]{}%
\providecommand \bibitemStop [0]{}%
\providecommand \bibitemNoStop [0]{.\EOS\space}%
\providecommand \EOS [0]{\spacefactor3000\relax}%
\providecommand \BibitemShut  [1]{\csname bibitem#1\endcsname}%
\let\auto@bib@innerbib\@empty
\bibitem [{\citenamefont {Aaij}\ \emph {et~al.}(2019)\citenamefont {Aaij} \emph
  {et~al.}}]{Aaij:2019vzc}%
  \BibitemOpen
  \bibfield  {author} {\bibinfo {author} {\bibfnamefont {Roel}\ \bibnamefont
  {Aaij}} \emph {et~al.} (\bibinfo {collaboration} {LHCb}),\ }\bibfield
  {title} {\enquote {\bibinfo {title} {{Observation of a narrow pentaquark
  state, $P_c(4312)^+$, and of two-peak structure of the $P_c(4450)^+$}},}\
  }\href {\doibase 10.1103/PhysRevLett.122.222001} {\bibfield  {journal}
  {\bibinfo  {journal} {Phys. Rev. Lett.}\ }\textbf {\bibinfo {volume} {122}},\
  \bibinfo {pages} {222001} (\bibinfo {year} {2019})},\ \Eprint
  {http://arxiv.org/abs/1904.03947} {arXiv:1904.03947 [hep-ex]} \BibitemShut
  {NoStop}%
\bibitem [{\citenamefont {Aaij}\ \emph {et~al.}(2015)\citenamefont {Aaij} \emph
  {et~al.}}]{Aaij:2015tga}%
  \BibitemOpen
  \bibfield  {author} {\bibinfo {author} {\bibfnamefont {Roel}\ \bibnamefont
  {Aaij}} \emph {et~al.} (\bibinfo {collaboration} {LHCb}),\ }\bibfield
  {title} {\enquote {\bibinfo {title} {{Observation of $J/\psi p$ Resonances
  Consistent with Pentaquark States in $\Lambda_b^0 \to J/\psi K^- p$
  Decays}},}\ }\href {\doibase 10.1103/PhysRevLett.115.072001} {\bibfield
  {journal} {\bibinfo  {journal} {Phys. Rev. Lett.}\ }\textbf {\bibinfo
  {volume} {115}},\ \bibinfo {pages} {072001} (\bibinfo {year} {2015})},\
  \Eprint {http://arxiv.org/abs/1507.03414} {arXiv:1507.03414 [hep-ex]}
  \BibitemShut {NoStop}%
\bibitem [{\citenamefont {Burns}(2015)}]{Burns:2015dwa}%
  \BibitemOpen
  \bibfield  {author} {\bibinfo {author} {\bibfnamefont {T.~J.}\ \bibnamefont
  {Burns}},\ }\bibfield  {title} {\enquote {\bibinfo {title} {{Phenomenology of
  P$_{c}$(4380)$^{+}$, P$_{c}$(4450)$^{+}$ and related states}},}\ }\href
  {\doibase 10.1140/epja/i2015-15152-6} {\bibfield  {journal} {\bibinfo
  {journal} {Eur. Phys. J. A}\ }\textbf {\bibinfo {volume} {51}},\ \bibinfo
  {pages} {152} (\bibinfo {year} {2015})},\ \Eprint
  {http://arxiv.org/abs/1509.02460} {arXiv:1509.02460 [hep-ph]} \BibitemShut
  {NoStop}%
\bibitem [{\citenamefont {Richard}(2016)}]{Richard:2016eis}%
  \BibitemOpen
  \bibfield  {author} {\bibinfo {author} {\bibfnamefont {Jean-Marc}\
  \bibnamefont {Richard}},\ }\bibfield  {title} {\enquote {\bibinfo {title}
  {{Exotic hadrons: review and perspectives}},}\ }\href {\doibase
  10.1007/s00601-016-1159-0} {\bibfield  {journal} {\bibinfo  {journal} {Few
  Body Syst.}\ }\textbf {\bibinfo {volume} {57}},\ \bibinfo {pages}
  {1185--1212} (\bibinfo {year} {2016})},\ \Eprint
  {http://arxiv.org/abs/1606.08593} {arXiv:1606.08593 [hep-ph]} \BibitemShut
  {NoStop}%
\bibitem [{\citenamefont {Lebed}\ \emph {et~al.}(2017)\citenamefont {Lebed},
  \citenamefont {Mitchell},\ and\ \citenamefont {Swanson}}]{Lebed:2016hpi}%
  \BibitemOpen
  \bibfield  {author} {\bibinfo {author} {\bibfnamefont {Richard~F.}\
  \bibnamefont {Lebed}}, \bibinfo {author} {\bibfnamefont {Ryan~E.}\
  \bibnamefont {Mitchell}}, \ and\ \bibinfo {author} {\bibfnamefont {Eric~S.}\
  \bibnamefont {Swanson}},\ }\bibfield  {title} {\enquote {\bibinfo {title}
  {{Heavy-Quark QCD Exotica}},}\ }\href {\doibase 10.1016/j.ppnp.2016.11.003}
  {\bibfield  {journal} {\bibinfo  {journal} {Prog. Part. Nucl. Phys.}\
  }\textbf {\bibinfo {volume} {93}},\ \bibinfo {pages} {143--194} (\bibinfo
  {year} {2017})},\ \Eprint {http://arxiv.org/abs/1610.04528} {arXiv:1610.04528
  [hep-ph]} \BibitemShut {NoStop}%
\bibitem [{\citenamefont {Esposito}\ \emph {et~al.}(2017)\citenamefont
  {Esposito}, \citenamefont {Pilloni},\ and\ \citenamefont
  {Polosa}}]{Esposito:2016noz}%
  \BibitemOpen
  \bibfield  {author} {\bibinfo {author} {\bibfnamefont {A.}~\bibnamefont
  {Esposito}}, \bibinfo {author} {\bibfnamefont {A.}~\bibnamefont {Pilloni}}, \
  and\ \bibinfo {author} {\bibfnamefont {A.~D.}\ \bibnamefont {Polosa}},\
  }\bibfield  {title} {\enquote {\bibinfo {title} {{Multiquark Resonances}},}\
  }\href {\doibase 10.1016/j.physrep.2016.11.002} {\bibfield  {journal}
  {\bibinfo  {journal} {Phys. Rept.}\ }\textbf {\bibinfo {volume} {668}},\
  \bibinfo {pages} {1--97} (\bibinfo {year} {2017})},\ \Eprint
  {http://arxiv.org/abs/1611.07920} {arXiv:1611.07920 [hep-ph]} \BibitemShut
  {NoStop}%
\bibitem [{\citenamefont {Olsen}\ \emph {et~al.}(2018)\citenamefont {Olsen},
  \citenamefont {Skwarnicki},\ and\ \citenamefont {Zieminska}}]{Olsen:2017bmm}%
  \BibitemOpen
  \bibfield  {author} {\bibinfo {author} {\bibfnamefont {Stephen~Lars}\
  \bibnamefont {Olsen}}, \bibinfo {author} {\bibfnamefont {Tomasz}\
  \bibnamefont {Skwarnicki}}, \ and\ \bibinfo {author} {\bibfnamefont {Daria}\
  \bibnamefont {Zieminska}},\ }\bibfield  {title} {\enquote {\bibinfo {title}
  {{Nonstandard heavy mesons and baryons: Experimental evidence}},}\ }\href
  {\doibase 10.1103/RevModPhys.90.015003} {\bibfield  {journal} {\bibinfo
  {journal} {Rev. Mod. Phys.}\ }\textbf {\bibinfo {volume} {90}},\ \bibinfo
  {pages} {015003} (\bibinfo {year} {2018})},\ \Eprint
  {http://arxiv.org/abs/1708.04012} {arXiv:1708.04012 [hep-ph]} \BibitemShut
  {NoStop}%
\bibitem [{\citenamefont {Guo}\ \emph {et~al.}(2018)\citenamefont {Guo},
  \citenamefont {Hanhart}, \citenamefont {Mei\ss{}ner}, \citenamefont {Wang},
  \citenamefont {Zhao},\ and\ \citenamefont {Zou}}]{Guo:2017jvc}%
  \BibitemOpen
  \bibfield  {author} {\bibinfo {author} {\bibfnamefont {Feng-Kun}\
  \bibnamefont {Guo}}, \bibinfo {author} {\bibfnamefont {Christoph}\
  \bibnamefont {Hanhart}}, \bibinfo {author} {\bibfnamefont {Ulf-G.}\
  \bibnamefont {Mei\ss{}ner}}, \bibinfo {author} {\bibfnamefont {Qian}\
  \bibnamefont {Wang}}, \bibinfo {author} {\bibfnamefont {Qiang}\ \bibnamefont
  {Zhao}}, \ and\ \bibinfo {author} {\bibfnamefont {Bing-Song}\ \bibnamefont
  {Zou}},\ }\bibfield  {title} {\enquote {\bibinfo {title} {{Hadronic
  molecules}},}\ }\href {\doibase 10.1103/RevModPhys.90.015004} {\bibfield
  {journal} {\bibinfo  {journal} {Rev. Mod. Phys.}\ }\textbf {\bibinfo {volume}
  {90}},\ \bibinfo {pages} {015004} (\bibinfo {year} {2018})},\ \Eprint
  {http://arxiv.org/abs/1705.00141} {arXiv:1705.00141 [hep-ph]} \BibitemShut
  {NoStop}%
\bibitem [{\citenamefont {Karliner}\ \emph {et~al.}(2018)\citenamefont
  {Karliner}, \citenamefont {Rosner},\ and\ \citenamefont
  {Skwarnicki}}]{Karliner:2017qhf}%
  \BibitemOpen
  \bibfield  {author} {\bibinfo {author} {\bibfnamefont {Marek}\ \bibnamefont
  {Karliner}}, \bibinfo {author} {\bibfnamefont {Jonathan~L.}\ \bibnamefont
  {Rosner}}, \ and\ \bibinfo {author} {\bibfnamefont {Tomasz}\ \bibnamefont
  {Skwarnicki}},\ }\bibfield  {title} {\enquote {\bibinfo {title} {{Multiquark
  States}},}\ }\href {\doibase 10.1146/annurev-nucl-101917-020902} {\bibfield
  {journal} {\bibinfo  {journal} {Ann. Rev. Nucl. Part. Sci.}\ }\textbf
  {\bibinfo {volume} {68}},\ \bibinfo {pages} {17--44} (\bibinfo {year}
  {2018})},\ \Eprint {http://arxiv.org/abs/1711.10626} {arXiv:1711.10626
  [hep-ph]} \BibitemShut {NoStop}%
\bibitem [{\citenamefont {Du}\ \emph {et~al.}(2021)\citenamefont {Du},
  \citenamefont {Baru}, \citenamefont {Guo}, \citenamefont {Hanhart},
  \citenamefont {Mei\ss{}ner}, \citenamefont {Oller},\ and\ \citenamefont
  {Wang}}]{Du:2021fmf}%
  \BibitemOpen
  \bibfield  {author} {\bibinfo {author} {\bibfnamefont {Meng-Lin}\
  \bibnamefont {Du}}, \bibinfo {author} {\bibfnamefont {Vadim}\ \bibnamefont
  {Baru}}, \bibinfo {author} {\bibfnamefont {Feng-Kun}\ \bibnamefont {Guo}},
  \bibinfo {author} {\bibfnamefont {Christoph}\ \bibnamefont {Hanhart}},
  \bibinfo {author} {\bibfnamefont {Ulf-G.}\ \bibnamefont {Mei\ss{}ner}},
  \bibinfo {author} {\bibfnamefont {Jos\'e~A.}\ \bibnamefont {Oller}}, \ and\
  \bibinfo {author} {\bibfnamefont {Qian}\ \bibnamefont {Wang}},\ }\bibfield
  {title} {\enquote {\bibinfo {title} {{Revisiting the nature of the $P_c$
  pentaquarks}},}\ }\href@noop {} {\  (\bibinfo {year} {2021})},\ \Eprint
  {http://arxiv.org/abs/2102.07159} {arXiv:2102.07159 [hep-ph]} \BibitemShut
  {NoStop}%
\bibitem [{\citenamefont {Liu}\ and\ \citenamefont
  {Zahed}(2017{\natexlab{a}})}]{Liu:2017xzo}%
  \BibitemOpen
  \bibfield  {author} {\bibinfo {author} {\bibfnamefont {Yizhuang}\
  \bibnamefont {Liu}}\ and\ \bibinfo {author} {\bibfnamefont {Ismail}\
  \bibnamefont {Zahed}},\ }\bibfield  {title} {\enquote {\bibinfo {title}
  {{Heavy Baryons and their Exotics from Instantons in Holographic QCD}},}\
  }\href {\doibase 10.1103/PhysRevD.95.116012} {\bibfield  {journal} {\bibinfo
  {journal} {Phys. Rev. D}\ }\textbf {\bibinfo {volume} {95}},\ \bibinfo
  {pages} {116012} (\bibinfo {year} {2017}{\natexlab{a}})},\ \Eprint
  {http://arxiv.org/abs/1704.03412} {arXiv:1704.03412 [hep-ph]} \BibitemShut
  {NoStop}%
\bibitem [{\citenamefont {Liu}\ and\ \citenamefont
  {Zahed}(2017{\natexlab{b}})}]{Liu:2017frj}%
  \BibitemOpen
  \bibfield  {author} {\bibinfo {author} {\bibfnamefont {Yizhuang}\
  \bibnamefont {Liu}}\ and\ \bibinfo {author} {\bibfnamefont {Ismail}\
  \bibnamefont {Zahed}},\ }\bibfield  {title} {\enquote {\bibinfo {title}
  {{Heavy and Strange Holographic Baryons}},}\ }\href {\doibase
  10.1103/PhysRevD.96.056027} {\bibfield  {journal} {\bibinfo  {journal} {Phys.
  Rev. D}\ }\textbf {\bibinfo {volume} {96}},\ \bibinfo {pages} {056027}
  (\bibinfo {year} {2017}{\natexlab{b}})},\ \Eprint
  {http://arxiv.org/abs/1705.01397} {arXiv:1705.01397 [hep-ph]} \BibitemShut
  {NoStop}%
\bibitem [{\citenamefont {Liu}\ \emph {et~al.}(2019)\citenamefont {Liu},
  \citenamefont {Pan}, \citenamefont {Peng}, \citenamefont
  {S\'anchez~S\'anchez}, \citenamefont {Geng}, \citenamefont {Hosaka},\ and\
  \citenamefont {Pavon~Valderrama}}]{Liu:2019tjn}%
  \BibitemOpen
  \bibfield  {author} {\bibinfo {author} {\bibfnamefont {Ming-Zhu}\
  \bibnamefont {Liu}}, \bibinfo {author} {\bibfnamefont {Ya-Wen}\ \bibnamefont
  {Pan}}, \bibinfo {author} {\bibfnamefont {Fang-Zheng}\ \bibnamefont {Peng}},
  \bibinfo {author} {\bibfnamefont {Mario}\ \bibnamefont
  {S\'anchez~S\'anchez}}, \bibinfo {author} {\bibfnamefont {Li-Sheng}\
  \bibnamefont {Geng}}, \bibinfo {author} {\bibfnamefont {Atsushi}\
  \bibnamefont {Hosaka}}, \ and\ \bibinfo {author} {\bibfnamefont {Manuel}\
  \bibnamefont {Pavon~Valderrama}},\ }\bibfield  {title} {\enquote {\bibinfo
  {title} {{Emergence of a complete heavy-quark spin symmetry multiplet: seven
  molecular pentaquarks in light of the latest LHCb analysis}},}\ }\href
  {\doibase 10.1103/PhysRevLett.122.242001} {\bibfield  {journal} {\bibinfo
  {journal} {Phys. Rev. Lett.}\ }\textbf {\bibinfo {volume} {122}},\ \bibinfo
  {pages} {242001} (\bibinfo {year} {2019})},\ \Eprint
  {http://arxiv.org/abs/1903.11560} {arXiv:1903.11560 [hep-ph]} \BibitemShut
  {NoStop}%
\bibitem [{\citenamefont {Xiao}\ \emph {et~al.}(2019)\citenamefont {Xiao},
  \citenamefont {Nieves},\ and\ \citenamefont {Oset}}]{Xiao:2019aya}%
  \BibitemOpen
  \bibfield  {author} {\bibinfo {author} {\bibfnamefont {C.~W.}\ \bibnamefont
  {Xiao}}, \bibinfo {author} {\bibfnamefont {J.}~\bibnamefont {Nieves}}, \ and\
  \bibinfo {author} {\bibfnamefont {E.}~\bibnamefont {Oset}},\ }\bibfield
  {title} {\enquote {\bibinfo {title} {{Heavy quark spin symmetric molecular
  states from ${\bar D}^{(*)}\Sigma_c^{(*)}$ and other coupled channels in the
  light of the recent LHCb pentaquarks}},}\ }\href {\doibase
  10.1103/PhysRevD.100.014021} {\bibfield  {journal} {\bibinfo  {journal}
  {Phys. Rev. D}\ }\textbf {\bibinfo {volume} {100}},\ \bibinfo {pages}
  {014021} (\bibinfo {year} {2019})},\ \Eprint
  {http://arxiv.org/abs/1904.01296} {arXiv:1904.01296 [hep-ph]} \BibitemShut
  {NoStop}%
\bibitem [{\citenamefont {Du}\ \emph {et~al.}(2020)\citenamefont {Du},
  \citenamefont {Baru}, \citenamefont {Guo}, \citenamefont {Hanhart},
  \citenamefont {Mei\ss{}ner}, \citenamefont {Oller},\ and\ \citenamefont
  {Wang}}]{Du:2019pij}%
  \BibitemOpen
  \bibfield  {author} {\bibinfo {author} {\bibfnamefont {Meng-Lin}\
  \bibnamefont {Du}}, \bibinfo {author} {\bibfnamefont {Vadim}\ \bibnamefont
  {Baru}}, \bibinfo {author} {\bibfnamefont {Feng-Kun}\ \bibnamefont {Guo}},
  \bibinfo {author} {\bibfnamefont {Christoph}\ \bibnamefont {Hanhart}},
  \bibinfo {author} {\bibfnamefont {Ulf-G}\ \bibnamefont {Mei\ss{}ner}},
  \bibinfo {author} {\bibfnamefont {Jos\'e~A.}\ \bibnamefont {Oller}}, \ and\
  \bibinfo {author} {\bibfnamefont {Qian}\ \bibnamefont {Wang}},\ }\bibfield
  {title} {\enquote {\bibinfo {title} {{Interpretation of the LHCb $P_c$ States
  as Hadronic Molecules and Hints of a Narrow $P_c(4380)$}},}\ }\href {\doibase
  10.1103/PhysRevLett.124.072001} {\bibfield  {journal} {\bibinfo  {journal}
  {Phys. Rev. Lett.}\ }\textbf {\bibinfo {volume} {124}},\ \bibinfo {pages}
  {072001} (\bibinfo {year} {2020})},\ \Eprint
  {http://arxiv.org/abs/1910.11846} {arXiv:1910.11846 [hep-ph]} \BibitemShut
  {NoStop}%
\bibitem [{\citenamefont {Yan}\ \emph {et~al.}(2021)\citenamefont {Yan},
  \citenamefont {Peng}, \citenamefont {S\'anchez},\ and\ \citenamefont
  {Valderrama}}]{Yan:2021nio}%
  \BibitemOpen
  \bibfield  {author} {\bibinfo {author} {\bibfnamefont {Mao-Jun}\ \bibnamefont
  {Yan}}, \bibinfo {author} {\bibfnamefont {Fang-Zheng}\ \bibnamefont {Peng}},
  \bibinfo {author} {\bibfnamefont {Mario~S\'anchez}\ \bibnamefont
  {S\'anchez}}, \ and\ \bibinfo {author} {\bibfnamefont {Manuel~Pavon}\
  \bibnamefont {Valderrama}},\ }\bibfield  {title} {\enquote {\bibinfo {title}
  {{Interpretations of the new LHCb $P_c(4337)^+$ pentaquark state}},}\
  }\href@noop {} {\  (\bibinfo {year} {2021})},\ \Eprint
  {http://arxiv.org/abs/2108.05306} {arXiv:2108.05306 [hep-ph]} \BibitemShut
  {NoStop}%
\bibitem [{\citenamefont {Aaij}\ \emph {et~al.}(2021)\citenamefont {Aaij} \emph
  {et~al.}}]{LHCb:2021chn}%
  \BibitemOpen
  \bibfield  {author} {\bibinfo {author} {\bibfnamefont {Roel}\ \bibnamefont
  {Aaij}} \emph {et~al.} (\bibinfo {collaboration} {LHCb}),\ }\bibfield
  {title} {\enquote {\bibinfo {title} {{Evidence for a new structure in the
  $J/\psi p$ and $J/\psi \bar{p}$ systems in $B_s^0 \to J/\psi p \bar{p}$
  decays}},}\ }\href@noop {} {\  (\bibinfo {year} {2021})},\ \Eprint
  {http://arxiv.org/abs/2108.04720} {arXiv:2108.04720 [hep-ex]} \BibitemShut
  {NoStop}%
\bibitem [{\citenamefont {Liu}\ \emph {et~al.}(2021)\citenamefont {Liu},
  \citenamefont {Nowak},\ and\ \citenamefont {Zahed}}]{Liu:2021tpq}%
  \BibitemOpen
  \bibfield  {author} {\bibinfo {author} {\bibfnamefont {Yizhuang}\
  \bibnamefont {Liu}}, \bibinfo {author} {\bibfnamefont {Maciej~A.}\
  \bibnamefont {Nowak}}, \ and\ \bibinfo {author} {\bibfnamefont {Ismail}\
  \bibnamefont {Zahed}},\ }\bibfield  {title} {\enquote {\bibinfo {title}
  {{Holographic charm and bottom pentaquarks I: Mass spectra with spin
  effects}},}\ }\href@noop {} {\  (\bibinfo {year} {2021})},\ \Eprint
  {http://arxiv.org/abs/2108.04334} {arXiv:2108.04334 [hep-ph]} \BibitemShut
  {NoStop}%
\bibitem [{\citenamefont {Li}(2017)}]{Li:2017dml}%
  \BibitemOpen
  \bibfield  {author} {\bibinfo {author} {\bibfnamefont {Si-wen}\ \bibnamefont
  {Li}},\ }\bibfield  {title} {\enquote {\bibinfo {title} {{Holographic
  heavy-baryons in the Witten-Sakai-Sugimoto model with the D0-D4
  background}},}\ }\href {\doibase 10.1103/PhysRevD.96.106018} {\bibfield
  {journal} {\bibinfo  {journal} {Phys. Rev. D}\ }\textbf {\bibinfo {volume}
  {96}},\ \bibinfo {pages} {106018} (\bibinfo {year} {2017})},\ \Eprint
  {http://arxiv.org/abs/1707.06439} {arXiv:1707.06439 [hep-th]} \BibitemShut
  {NoStop}%
\bibitem [{\citenamefont {Fujii}\ and\ \citenamefont
  {Hosaka}(2020)}]{Fujii:2020jre}%
  \BibitemOpen
  \bibfield  {author} {\bibinfo {author} {\bibfnamefont {Daisuke}\ \bibnamefont
  {Fujii}}\ and\ \bibinfo {author} {\bibfnamefont {Atsushi}\ \bibnamefont
  {Hosaka}},\ }\bibfield  {title} {\enquote {\bibinfo {title} {{Heavy baryons
  in holographic QCD with higher dimensional degrees of freedom}},}\ }\href
  {\doibase 10.1103/PhysRevD.101.126008} {\bibfield  {journal} {\bibinfo
  {journal} {Phys. Rev. D}\ }\textbf {\bibinfo {volume} {101}},\ \bibinfo
  {pages} {126008} (\bibinfo {year} {2020})},\ \Eprint
  {http://arxiv.org/abs/2003.13415} {arXiv:2003.13415 [hep-ph]} \BibitemShut
  {NoStop}%
\bibitem [{\citenamefont {Sakai}\ and\ \citenamefont
  {Sugimoto}(2005)}]{Sakai:2004cn}%
  \BibitemOpen
  \bibfield  {author} {\bibinfo {author} {\bibfnamefont {Tadakatsu}\
  \bibnamefont {Sakai}}\ and\ \bibinfo {author} {\bibfnamefont {Shigeki}\
  \bibnamefont {Sugimoto}},\ }\bibfield  {title} {\enquote {\bibinfo {title}
  {{Low energy hadron physics in holographic QCD}},}\ }\href {\doibase
  10.1143/PTP.113.843} {\bibfield  {journal} {\bibinfo  {journal} {Prog. Theor.
  Phys.}\ }\textbf {\bibinfo {volume} {113}},\ \bibinfo {pages} {843--882}
  (\bibinfo {year} {2005})},\ \Eprint {http://arxiv.org/abs/hep-th/0412141}
  {arXiv:hep-th/0412141} \BibitemShut {NoStop}%
\bibitem [{\citenamefont {Liu}\ and\ \citenamefont
  {Zahed}(2017{\natexlab{c}})}]{Liu:2016iqo}%
  \BibitemOpen
  \bibfield  {author} {\bibinfo {author} {\bibfnamefont {Yizhuang}\
  \bibnamefont {Liu}}\ and\ \bibinfo {author} {\bibfnamefont {Ismail}\
  \bibnamefont {Zahed}},\ }\bibfield  {title} {\enquote {\bibinfo {title}
  {{Holographic Heavy-Light Chiral Effective Action}},}\ }\href {\doibase
  10.1103/PhysRevD.95.056022} {\bibfield  {journal} {\bibinfo  {journal} {Phys.
  Rev. D}\ }\textbf {\bibinfo {volume} {95}},\ \bibinfo {pages} {056022}
  (\bibinfo {year} {2017}{\natexlab{c}})},\ \Eprint
  {http://arxiv.org/abs/1611.03757} {arXiv:1611.03757 [hep-ph]} \BibitemShut
  {NoStop}%
\bibitem [{\citenamefont {Eides}\ \emph {et~al.}(2018)\citenamefont {Eides},
  \citenamefont {Petrov},\ and\ \citenamefont {Polyakov}}]{Eides:2017xnt}%
  \BibitemOpen
  \bibfield  {author} {\bibinfo {author} {\bibfnamefont {Michael~I.}\
  \bibnamefont {Eides}}, \bibinfo {author} {\bibfnamefont {Victor~Yu.}\
  \bibnamefont {Petrov}}, \ and\ \bibinfo {author} {\bibfnamefont {Maxim~V.}\
  \bibnamefont {Polyakov}},\ }\bibfield  {title} {\enquote {\bibinfo {title}
  {{Pentaquarks with hidden charm as hadroquarkonia}},}\ }\href {\doibase
  10.1140/epjc/s10052-018-5530-9} {\bibfield  {journal} {\bibinfo  {journal}
  {Eur. Phys. J. C}\ }\textbf {\bibinfo {volume} {78}},\ \bibinfo {pages} {36}
  (\bibinfo {year} {2018})},\ \Eprint {http://arxiv.org/abs/1709.09523}
  {arXiv:1709.09523 [hep-ph]} \BibitemShut {NoStop}%
\bibitem [{\citenamefont {Lin}\ and\ \citenamefont {Zou}(2019)}]{Lin:2019qiv}%
  \BibitemOpen
  \bibfield  {author} {\bibinfo {author} {\bibfnamefont {Yong-Hui}\
  \bibnamefont {Lin}}\ and\ \bibinfo {author} {\bibfnamefont {Bing-Song}\
  \bibnamefont {Zou}},\ }\bibfield  {title} {\enquote {\bibinfo {title}
  {{Strong decays of the latest LHCb pentaquark candidates in hadronic molecule
  pictures}},}\ }\href {\doibase 10.1103/PhysRevD.100.056005} {\bibfield
  {journal} {\bibinfo  {journal} {Phys. Rev. D}\ }\textbf {\bibinfo {volume}
  {100}},\ \bibinfo {pages} {056005} (\bibinfo {year} {2019})},\ \Eprint
  {http://arxiv.org/abs/1908.05309} {arXiv:1908.05309 [hep-ph]} \BibitemShut
  {NoStop}%
\bibitem [{\citenamefont {Hashimoto}\ \emph {et~al.}(2008)\citenamefont
  {Hashimoto}, \citenamefont {Sakai},\ and\ \citenamefont
  {Sugimoto}}]{Hashimoto:2008zw}%
  \BibitemOpen
  \bibfield  {author} {\bibinfo {author} {\bibfnamefont {Koji}\ \bibnamefont
  {Hashimoto}}, \bibinfo {author} {\bibfnamefont {Tadakatsu}\ \bibnamefont
  {Sakai}}, \ and\ \bibinfo {author} {\bibfnamefont {Shigeki}\ \bibnamefont
  {Sugimoto}},\ }\bibfield  {title} {\enquote {\bibinfo {title} {{Holographic
  Baryons: Static Properties and Form Factors from Gauge/String Duality}},}\
  }\href {\doibase 10.1143/PTP.120.1093} {\bibfield  {journal} {\bibinfo
  {journal} {Prog. Theor. Phys.}\ }\textbf {\bibinfo {volume} {120}},\ \bibinfo
  {pages} {1093--1137} (\bibinfo {year} {2008})},\ \Eprint
  {http://arxiv.org/abs/0806.3122} {arXiv:0806.3122 [hep-th]} \BibitemShut
  {NoStop}%
\bibitem [{\citenamefont {Eides}\ and\ \citenamefont
  {Petrov}(2018)}]{Eides:2018lqg}%
  \BibitemOpen
  \bibfield  {author} {\bibinfo {author} {\bibfnamefont {Michael~I.}\
  \bibnamefont {Eides}}\ and\ \bibinfo {author} {\bibfnamefont {Victor~Yu.}\
  \bibnamefont {Petrov}},\ }\bibfield  {title} {\enquote {\bibinfo {title}
  {{Decays of pentaquarks in hadrocharmonium and molecular scenarios}},}\
  }\href {\doibase 10.1103/PhysRevD.98.114037} {\bibfield  {journal} {\bibinfo
  {journal} {Phys. Rev. D}\ }\textbf {\bibinfo {volume} {98}},\ \bibinfo
  {pages} {114037} (\bibinfo {year} {2018})},\ \Eprint
  {http://arxiv.org/abs/1811.01691} {arXiv:1811.01691 [hep-ph]} \BibitemShut
  {NoStop}%
\bibitem [{\citenamefont {Nowak}\ \emph {et~al.}(1993)\citenamefont {Nowak},
  \citenamefont {Rho},\ and\ \citenamefont {Zahed}}]{Nowak:1992um}%
  \BibitemOpen
  \bibfield  {author} {\bibinfo {author} {\bibfnamefont {Maciej~A.}\
  \bibnamefont {Nowak}}, \bibinfo {author} {\bibfnamefont {Mannque}\
  \bibnamefont {Rho}}, \ and\ \bibinfo {author} {\bibfnamefont
  {I.}~\bibnamefont {Zahed}},\ }\bibfield  {title} {\enquote {\bibinfo {title}
  {{Chiral effective action with heavy quark symmetry}},}\ }\href {\doibase
  10.1103/PhysRevD.48.4370} {\bibfield  {journal} {\bibinfo  {journal} {Phys.
  Rev. D}\ }\textbf {\bibinfo {volume} {48}},\ \bibinfo {pages} {4370--4374}
  (\bibinfo {year} {1993})},\ \Eprint {http://arxiv.org/abs/hep-ph/9209272}
  {arXiv:hep-ph/9209272} \BibitemShut {NoStop}%
\bibitem [{\citenamefont {Bardeen}\ \emph {et~al.}(2003)\citenamefont
  {Bardeen}, \citenamefont {Eichten},\ and\ \citenamefont
  {Hill}}]{Bardeen:2003kt}%
  \BibitemOpen
  \bibfield  {author} {\bibinfo {author} {\bibfnamefont {William~A.}\
  \bibnamefont {Bardeen}}, \bibinfo {author} {\bibfnamefont {Estia~J.}\
  \bibnamefont {Eichten}}, \ and\ \bibinfo {author} {\bibfnamefont
  {Christopher~T.}\ \bibnamefont {Hill}},\ }\bibfield  {title} {\enquote
  {\bibinfo {title} {{Chiral multiplets of heavy - light mesons}},}\ }\href
  {\doibase 10.1103/PhysRevD.68.054024} {\bibfield  {journal} {\bibinfo
  {journal} {Phys. Rev. D}\ }\textbf {\bibinfo {volume} {68}},\ \bibinfo
  {pages} {054024} (\bibinfo {year} {2003})},\ \Eprint
  {http://arxiv.org/abs/hep-ph/0305049} {arXiv:hep-ph/0305049} \BibitemShut
  {NoStop}%
\bibitem [{\citenamefont {Liu}\ and\ \citenamefont
  {Zahed}(2016)}]{Liu:2016kqx}%
  \BibitemOpen
  \bibfield  {author} {\bibinfo {author} {\bibfnamefont {Yizhuang}\
  \bibnamefont {Liu}}\ and\ \bibinfo {author} {\bibfnamefont {Ismail}\
  \bibnamefont {Zahed}},\ }\bibfield  {title} {\enquote {\bibinfo {title}
  {{Heavy Exotic Molecules with Charm and Bottom}},}\ }\href {\doibase
  10.1016/j.physletb.2016.09.045} {\bibfield  {journal} {\bibinfo  {journal}
  {Phys. Lett. B}\ }\textbf {\bibinfo {volume} {762}},\ \bibinfo {pages}
  {362--370} (\bibinfo {year} {2016})},\ \Eprint
  {http://arxiv.org/abs/1608.06535} {arXiv:1608.06535 [hep-ph]} \BibitemShut
  {NoStop}%
\bibitem [{\citenamefont {Wang}\ \emph {et~al.}(2015)\citenamefont {Wang},
  \citenamefont {Liu},\ and\ \citenamefont {Zhao}}]{Wang:2015jsa}%
  \BibitemOpen
  \bibfield  {author} {\bibinfo {author} {\bibfnamefont {Qian}\ \bibnamefont
  {Wang}}, \bibinfo {author} {\bibfnamefont {Xiao-Hai}\ \bibnamefont {Liu}}, \
  and\ \bibinfo {author} {\bibfnamefont {Qiang}\ \bibnamefont {Zhao}},\
  }\bibfield  {title} {\enquote {\bibinfo {title} {{Photoproduction of hidden
  charm pentaquark states $P_c^+(4380)$ and $P_c^+(4450)$}},}\ }\href {\doibase
  10.1103/PhysRevD.92.034022} {\bibfield  {journal} {\bibinfo  {journal} {Phys.
  Rev. D}\ }\textbf {\bibinfo {volume} {92}},\ \bibinfo {pages} {034022}
  (\bibinfo {year} {2015})},\ \Eprint {http://arxiv.org/abs/1508.00339}
  {arXiv:1508.00339 [hep-ph]} \BibitemShut {NoStop}%
\bibitem [{\citenamefont {Kubarovsky}\ and\ \citenamefont
  {Voloshin}(2015)}]{Kubarovsky:2015aaa}%
  \BibitemOpen
  \bibfield  {author} {\bibinfo {author} {\bibfnamefont {V.}~\bibnamefont
  {Kubarovsky}}\ and\ \bibinfo {author} {\bibfnamefont {M.~B.}\ \bibnamefont
  {Voloshin}},\ }\bibfield  {title} {\enquote {\bibinfo {title} {{Formation of
  hidden-charm pentaquarks in photon-nucleon collisions}},}\ }\href {\doibase
  10.1103/PhysRevD.92.031502} {\bibfield  {journal} {\bibinfo  {journal} {Phys.
  Rev. D}\ }\textbf {\bibinfo {volume} {92}},\ \bibinfo {pages} {031502}
  (\bibinfo {year} {2015})},\ \Eprint {http://arxiv.org/abs/1508.00888}
  {arXiv:1508.00888 [hep-ph]} \BibitemShut {NoStop}%
\bibitem [{\citenamefont {Karliner}\ and\ \citenamefont
  {Rosner}(2016)}]{Karliner:2015voa}%
  \BibitemOpen
  \bibfield  {author} {\bibinfo {author} {\bibfnamefont {Marek}\ \bibnamefont
  {Karliner}}\ and\ \bibinfo {author} {\bibfnamefont {Jonathan~L.}\
  \bibnamefont {Rosner}},\ }\bibfield  {title} {\enquote {\bibinfo {title}
  {{Photoproduction of Exotic Baryon Resonances}},}\ }\href {\doibase
  10.1016/j.physletb.2015.11.068} {\bibfield  {journal} {\bibinfo  {journal}
  {Phys. Lett. B}\ }\textbf {\bibinfo {volume} {752}},\ \bibinfo {pages}
  {329--332} (\bibinfo {year} {2016})},\ \Eprint
  {http://arxiv.org/abs/1508.01496} {arXiv:1508.01496 [hep-ph]} \BibitemShut
  {NoStop}%
\bibitem [{\citenamefont {Meziani}\ and\ \citenamefont
  {Joosten}(2020)}]{Meziani:2020oks}%
  \BibitemOpen
  \bibfield  {author} {\bibinfo {author} {\bibfnamefont {Zein-Eddine}\
  \bibnamefont {Meziani}}\ and\ \bibinfo {author} {\bibfnamefont {Sylvester}\
  \bibnamefont {Joosten}},\ }\bibfield  {title} {\enquote {\bibinfo {title}
  {{Origin of the Proton Mass? Heavy Quarkonium Production at Threshold from
  Jefferson Lab to an Electron Ion Collider}},}\ }in\ \href {\doibase
  10.1142/9789811214950_0048} {\emph {\bibinfo {booktitle} {{Probing Nucleons
  and Nuclei in High Energy Collisions}: {Dedicated to the Physics of the
  Electron Ion Collider}}}}\ (\bibinfo {year} {2020})\BibitemShut {NoStop}%
\end{thebibliography}%

\end{document}